\documentclass[epj,nopacs]{svjour}
\usepackage{microtype}
\usepackage{amsfonts,amsmath,amssymb,esint,array,dsfont,stmaryrd}
\usepackage{graphicx}
\usepackage{cite}
\usepackage[pdfstartview=FitH,colorlinks=true,linkcolor=blue,citecolor=blue,urlcolor=blue]{hyperref}

\begin{document}

\title{Pump-and-probe optical transmission phase shift as a quantitative probe of the Bogoliubov dispersion relation in a nonlinear channel waveguide}
\titlerunning{Pump-and-probe optical transmission phase shift as a quantitative probe of the{\dots}}

\author{
Pierre-\'Elie Larr\'e\inst{1}\thanks{\email{\href{mailto:pierre-elie.larre@lkb.upmc.fr}{pierre-elie.larre@lkb.upmc.fr}}} \and
Stefano Biasi\inst{2} \and
Fernando Ramiro-Manzano\inst{2} \and
Lorenzo Pavesi\inst{2} \and
Iacopo Carusotto\inst{3}\thanks{\email{\href{mailto:iacopo.carusotto@unitn.it}{iacopo.carusotto@unitn.it}}}
}
\authorrunning{Pierre-\'Elie Larr\'e et al.}

\institute{
Laboratoire Kastler-Brossel, UPMC (Sorbonne Universit\'e), CNRS, ENS (Universit\'e de Recherche PSL), Coll\`ege de France, 4 Place Jussieu, 75005 Paris, France \and
Laboratorio di Nanoscienze, Universit\`a degli Studi di Trento, CNR, INFM, Via Sommarive 14, 38123 Povo (TN), Italia \and
BEC Center, Universit\`a degli Studi di Trento, CNR, INO, Via Sommarive 14, 38123 Povo (TN), Italia
}

\date{\today}

\abstract{
We theoretically investigate the dispersion relation of small-amplitude optical waves superimposing upon a beam of polarized monochromatic light propagating along a single-mode channel waveguide characterized by an instantaneous and spatially local Kerr nonlinearity. These small luminous fluctuations propagate along the waveguide as Bogoliubov elementary excitations on top of a one-dimensional dilute Bose quantum fluid evolve in time. They consequently display a strongly renormalized dispersion law, of Bogoliubov type. Analytical and numerical results are found in both the absence and the presence of one- and two-photon losses. Silicon and silicon-nitride waveguides are used as examples. We finally propose an experiment to measure this Bogoliubov dispersion relation, based on a stimulated four-wave mixing and interference spectroscopy techniques.
%
}

\maketitle

\section{Introduction}
\label{Sec:Introduction}

The dynamics of small-amplitude perturbations on top of a weakly interacting Bose quantum fluid may be described within the framework of Bogoliubov's theory \cite{Castin2001, Pethick2002, Pitaevskii2016}: The elementary excitations of the fluid are collective bosonic excitations whose direct- and reciprocal-space profiles may be obtained by linearizing the Heisenberg equation of the system around the equilibrium state. In the broadest sense of the term, the Bogoliubov dispersion relation is the energy-momentum law of these Bogoliubov fluctuations. This physical quantity is conceptually important in physics. In particular, it is at the heart of the physics of Bose-Einstein condensation and superfluidity \cite{Castin2001, Pethick2002, Pitaevskii2016}.

The first most prominent application of the Bogoliubov theory of elementary excitations was formulated by Bogoliubov himself to qualitatively explain the superfluid behavior of strongly interacting quantum fluids like liquid helium \cite{Bogoliubov1947}. In this dense system, the elementary-excitation dispersion relation displays a phonon and roton behavior and was measured by means of experiments based on cold-neutron scattering \cite{Palevsky1958, Yarnell1959}. The quantitative experimental verification of the theory came later on with the realization of Bose-Einstein condensates in ultracold vapors of weakly interacting atoms \cite{Anderson1995, Davis1995}. In these dilute systems, the Bogoliubov dispersion relation presents a phonon and free-particle behavior and was experimentally obtained thanks to experiments based on two-photon Bragg scattering \cite{Stenger1999, StamperKurn1999, Vogels2002, Steinhauer2002, Ozeri2002, Katz2002, Steinhauer2003, Ozeri2005}.

In the last few years, the physics of quantum fluids extended to nonlinear photonics, embracing a novel class of systems, the so-called quantum fluids of light \cite{Carusotto2013}: In the presence (i) of a significant spatial confinement, and/or of a weak diffraction and a strong degree of monochromaticity, and (ii) of a substantial Kerr optical nonlinearity, light and matter may combine to generate photonlike particles that, differently from vacuum photons, are characterized by sizeable (i) effective masses and (ii) mutual interactions; in this case, a many-photon system may behave collectively as a quantum fluid of matter, with novel features stemming from its intrinsically nonequilibrium nature.

While the connection between hydrodynamics and nonlinear and laser optics started being exploited a few decades ago \cite{Staliunas2003}, the field of quantum fluids of light received a further boost with the advent of semiconductor microcavities in the regime of strong light-matter coupling \cite{Weisbuch1992}. These systems exhibit exciton polaritons \cite{Carusotto2013} as weakly interacting bosonic quasiparticles that, most remarkably, were experimentally demonstrated to condense and display long-range spatial coherence at cryogenic temperatures \cite{Kasprzak2006, Balili2007, Deng2007, Deng2010}. As they are naturally subject to driving and dissipation, these dilute cavity exciton-polariton condensates are characterized by Bogoliubov dispersion relations with several novel out-of-equilibrium features that were widely investigated at both the theoretical \cite{Shelykh2005, Szymanska2006, Wouters2007, Wouters2008, Wouters2009, Byrnes2012, Larre2012, Larre2013} and the experimental \cite{Kohnle2011, Kohnle2012} level.

An optical platform alternative to these microcavity devices---and which presently attracts a growing interest within the quantum-fluid-of-light community (see, e.g., References~\cite{Leboeuf2010, Elazar2013, Carusotto2014, Larre2015a, Vocke2015, Larre2015b, Karpov2015, Larre2016a, Vocke2016, Larre2016b, Gullans2016, Noh2016} to cite a few recent works)---consists in a paraxial beam of quasimonochromatic light propagating in a cavityless, bulk, Kerr medium. In contrast to the intrinsically driven-dissipative dynamics of the field in laser media and/or cavities \cite{Carusotto2013, Staliunas2003}, in such propagating geometries, it is well known that the complex amplitude of the optical field is a slowly varying function of space and time whose propagation is ruled by a nonlinear Schr\"odinger equation \cite{Boyd1992, Agrawal1995, Rosanov2002} mathematically identical, in the absence of losses, to the conservative Gross-Pitaevskii equation of dilute atomic Bose-Einstein condensates \cite{Castin2001, Pethick2002, Pitaevskii2016} after exchanging the roles played by the time parameter and the propagation coordinate (see, e.g., References~\cite{Bortolozzo2009, Michel2011, Turitsyna2013, Perego2016, Tarasov2016, Kumar2017} to cite other recent works). This nonlinear propagation equation has been extensively used to describe many interesting nonlinear-optics dynamical phenomena such as the formation and the evolution of solitons \cite{Haus1991, Dunlop1998}, of small-amplitude waves \cite{Schopf1991, Genoud2010}, and of modulation instabilities \cite{Alexeeva1999}. In this latter case in particular, the wave equation predicts a modulation-instability spectrum \cite{Boyd1992, Agrawal1995, Rosanov2002} formally analogous to the Bogoliubov dispersion relation of a quantum gas of weakly attractive atom bosons \cite{Castin2001, Pethick2002, Pitaevskii2016}.

While this spectrum has been widely used to obtain the modulation-instability gain in nonlinear-fiber optics \cite{Agrawal1995, Zakharov2009, Turitsyn2010, Conforti2014}, so far, no special attention has been devoted to its study in its own right---in particular in terms of \textsl{dispersion relation of the elementary excitations of a propagating quantum fluid of light}---and to its measurement, exception made for the preliminary experimental work \cite{Vocke2015}, which was based on the nonlocal thermal optical nonlinearity of a liquid medium. In this article, we push this research line forward by theoretically investigating the Bogoliubov dispersion relation of a fluid of light propagating along a nonlinear single-mode channel \cite{Paschotta2008} waveguide and by proposing an experimental setup to measure it. Because it is nonperturbatively coupled to its four-wave-mixed partner, a weak-power probe on top of a strong-power pump acquires a dispersion relation with peculiar collective features such as a sound-like shape at low detuning. In contrast to Reference~\cite{Vocke2015} where the effective mass of the photons originates from paraxial diffraction in the plane transverse to the propagation axis, and then where the transverse-plane branch of the Bogoliubov law is considered, in our one-dimensional guiding geometry, we focus on the temporal branch of the dispersion because the effective mass originates from the second-order dispersion properties of the material, as recently reviewed by two of us in References~\cite{Larre2015b, Larre2016a}. Additionally, by considering realistic materials for the waveguide, and in particular the photon losses that characterize them, we quantize the conditions needed to extract the Bogoliubov dispersion relation of the propagating fluid of light, based on the experimental setup that we propose to measure it.

The paper is organized as follows. To begin with, we present in Section~\ref{Sec:Model} the considered theoretical model, fully accounting for one- and two-photon losses. We then solve it in Section~\ref{Sec:MonochromaticBeam} in the ideal case of a beam of monochromatic light. In Section~\ref{Sec:BogoliubovDispersionRelation}, we study both analytically (partly on the basis of Appendix \ref{App:AdiabaticTheoremForZDependentPropagatingOpticalSystems}) and numerically the dispersion relation of small-amplitude fluctuations on top of the previously-found monochromatic solution, within the framework of Bogoliubov's theory. Realistic examples of waveguides fabricated within the silicon-photonics technology \cite{Leuthold2010, Vivien2013} are reported. We propose in Section~\ref{Sec:Experiment} a pump-and-probe experiment \cite{Biasi2016} to measure this Bogoliubov-type dispersion relation, based on interference spectroscopy techniques. Finally, we sum up our results in Section~\ref{Sec:Conclusion}.

\section{Model}
\label{Sec:Model}

In this section, we introduce the theoretical model investigated in this work, of light propagation along a nonlinear and realistically lossy single-mode channel waveguide. Section \ref{SubSec:DissipativeGrossPitaevskiiEquation} is devoted to the presentation of the corresponding wave equation, of dissipative Gross-Pitaevskii type, and Section~\ref{SubSec:MadelungSFormulation} puts on its modulus-phase formulation within Madelung's approach of wave mechanics.

\subsection{Dissipative Gross-Pitaevskii equation}
\label{SubSec:DissipativeGrossPitaevskiiEquation}

We consider the propagation in the positive-$z$ direction of a spectrally narrow beam of linearly polarized (e.g., along the $x$ axis) light of central angular frequency $\mathrm{\Omega}$ and propagation constant $\beta_{0}=\beta(\omega=\mathrm{\Omega})$ along a single-mode channel waveguide of dispersion law $\beta(\omega)$ and whose core displays an instantaneous and spatially local Kerr nonlinearity as well as one-photon losses and two-photon absorption at $\mathrm{\Omega}$. In this case, the amplitude $A(t,z)$ of the light wave's complex electric field \cite{Agrawal1995}
\begin{equation}
\label{Eq:ComplexElectricField}
F(x,y)\,A(t,z)\,e^{-i\mathrm{\Omega} t}\,e^{i\beta_{0}z}
\end{equation}
---where $F(x,y)$ denotes the normalized-to-unity transverse distribution of the waveguide's fundamental mode---is a slowly varying scalar function of the time parameter $t$ and of the propagation coordinate $z$ which satisfies the generalized nonlinear Schr\"odinger equation \cite{Agrawal1995}
\begin{equation}
\label{Eq:GrossPitaevskiiEquation}
i\,\frac{\partial A}{\partial z}=\frac{\beta_{2}}{2}\,\frac{\partial^{2}A}{\partial t^{2}}-\gamma\,|A|^{2}\,A-\frac{i}{2}\,(\alpha_{0}+\alpha_{2}\,|A|^{2})\,A,
\end{equation}
written here in the retarded frame, that is, the frame moving at the group velocity $1/\beta_{1}=1/[d\beta(\omega=\mathrm{\Omega})/d\omega]$ of the electric wave. This propagation equation makes use of the standard notations of nonlinear-fiber optics \cite{Agrawal1995}. In particular, $\beta_{2}=d^{2}\beta(\omega=\mathrm{\Omega})/d\omega^{2}\gtrless0$ denotes the group-velocity-dispersion parameter, $\gamma\gtrless0$ is the Kerr-nonlinearity coefficient, $\alpha_{0}\geqslant0$ is a linear coefficient describing one-photon propagation losses, and $\alpha_{2}\geqslant0$ is a nonlinear parameter describing two-photon absorption losses. These parameters are evaluated at the operating angular frequency $\mathrm{\Omega}$ and are expressed in terms of the vacuum speed of light $c$, the Kerr index $n_{2}$ (expressed here in $\mathrm{m}^{2}\cdot\mathrm{V}^{-2}$), the effective area $A_{\mathrm{eff}}=1/\iint_{-\infty}^{\infty}dx\,dy\,|F(x,y)|^{4}$ of the transverse mode, and the first- and third-order electric-susceptibility tensors $\chi^{(1)}$ and $\chi^{(3)}$ as \cite{Agrawal1995}
\begin{align}
\notag
\begin{bmatrix}
\gamma=(\mathrm{\Omega}/c)\,n_{2}/A_{\mathrm{eff}} \\
\alpha_{0} \\
\alpha_{2}
\end{bmatrix}
&\left.=\frac{\mathrm{\Omega}}{c}\,\frac{1}{1+\frac{1}{2}\,\mathrm{Re}(\chi_{x,x}^{(1)})}\right. \\
\label{Eq:OpticalParameters}
&\left.\hphantom{=}\times
\begin{bmatrix}
\frac{3}{8}\,\mathrm{Re}(\chi_{x,x,x,x}^{(3)})/A_{\mathrm{eff}} \vspace{1mm} \\
\mathrm{Im}(\chi_{x,x}^{(1)}) \vspace{1mm} \\
\frac{3}{4}\,\mathrm{Im}(\chi_{x,x,x,x}^{(3)})/A_{\mathrm{eff}}
\end{bmatrix}
.\right.
\end{align}
Higher-order dispersion terms \cite{Agrawal1995} are not included in the model and free-carrier absorption may be neglected (see, e.g., References~\cite{Ong2014, Borghi2015}). Equation \eqref{Eq:GrossPitaevskiiEquation}, which is of complex Ginzburg-Landau type, is nothing but the usual wave equation of nonlinear-fiber optics in the realistic case where one- and two-photon losses occur at $\mathrm{\Omega}$ \cite{Agrawal1995}. It is used to model many nonlinearity effects in one-dimensional optical waveguides, including, e.g., four-wave mixing, self-phase modulation, stimulated Raman scattering, or the formation of temporal solitons.

Most particularly, aside from the loss terms proportional to $\alpha_{0}$ and $\alpha_{2}$, it is formally analogous to the Gross-Pitaevskii equation of quasi-one-dimensional dilute atomic Bose-Einstein condensates \cite{Castin2001, Pethick2002, Pitaevskii2016}---hence the title of the present section---, generally used to describe nonlinear phenomena in one-dimensional atomic Bose quantum fluids such as, e.g., the formation of spatial solitons and shock waves, chaos effects, or nonlinear-tunneling superfluidlike phenomena and matter-wave Anderson localization in the presence of inhomogeneities. The mathematical analogy is here based on the five following points.
\begin{enumerate}
\item The time parameter and the position along the propagation axis, $t$ and $z$, play exchanged roles as spatial and temporal coordinates.
\item The complex amplitude of the electric field, $A(t,z)$, corresponds to the macroscopic single-particle wavefunction of the quasi-one-dimensional atomic condensate.
\item The opposite of the inverse of the second-order dispersion parameter, $-1/\beta_{2}$, is the analog of the atom mass.
\item The opposite of the optical-nonlinearity parameter, $-\gamma$, corresponds to the atom-atom interaction constant in the zero-range-pseudopotential approximation.
\item As a boundary condition, the temporal profile of the incident beam of light, given by $A(t,z=0)$, determines the initial condition on the solution of Equation~\eqref{Eq:GrossPitaevskiiEquation}, of first order in the timelike parameter $z$.
\end{enumerate}
Accordingly, in what follows, we shall often employ the language as well as mathematical techniques (mostly) specific to quantum hydrodynamics. For example, we will sometimes speak of ``fluid of light'' in place of ``beam of light,'' following the terminology used in the introductory Section~\ref{Sec:Introduction}, and in particular, we will use Bogoliubov's theory to describe the dispersion relation of the elementary excitations of the propagating beam. As far as we know, this has never been studied theoretically in its own right, neither within the nonlinear-optics community nor within the mathematical-physics one, be it in the nonlinear Schr\"odinger framework or the complex Ginzburg-Landau one.

\subsection{Madelung's formulation}
\label{SubSec:MadelungSFormulation}

Working within Madelung's formulation \cite{Petrov2003, Petrov2004} of the dissipative Gross-Pitaevskii-type equation \eqref{Eq:GrossPitaevskiiEquation} helps the analytical analysis of the here-investigated problem. To do so, one begins with writing the unknown of Equation~\eqref{Eq:GrossPitaevskiiEquation} in the following way:
\begin{equation}
\label{Eq:MadelungRepresentation}
A(t,z)=\sqrt{\rho(t,z)}\,e^{i\theta(t,z)},
\end{equation}
where $\rho(t,z)$ and $\theta(t,z)$ are real fields which physically correspond to the instantaneous, local power $P(t,z)=\frac{1}{2}\,c\,\varepsilon_{0}\,n_{0}\,\rho(t,z)$ ($\varepsilon_{0}$ is the vacuum permittivity and $n_{0}$ is the effective refractive index of the propagating mode at $\mathrm{\Omega}$) and to the instantaneous, local phase of the beam of light. Substituting the transformation \eqref{Eq:MadelungRepresentation} into Equation~\eqref{Eq:GrossPitaevskiiEquation} and separating the imaginary and real parts, we respectively get the following coupled equations for $\rho(t,z)$ and $\theta(t,z)$:
\begin{subequations}
\label{Eq:EulerEquations}
\begin{align}
\label{Eq:EulerEquations-a}
\frac{\partial\rho}{\partial z}&=\beta_{2}\,\frac{\partial}{\partial t}\bigg(\rho\,\frac{\partial\theta}{\partial t}\bigg)-(\alpha_{0}+\alpha_{2}\,\rho)\,\rho, \\
\label{Eq:EulerEquations-b}
\frac{\partial\theta}{\partial z}&=-\frac{\beta_{2}}{2\,\sqrt{\rho}}\,\frac{\partial^{2}\sqrt{\rho}}{\partial t^{2}}+\frac{\beta_{2}}{2}\,\bigg(\frac{\partial\theta}{\partial t}\bigg)^{2}+\gamma\,\rho.
\end{align}
\end{subequations}

Within the $t\longleftrightarrow z$ mapping discussed previously, Equations~\eqref{Eq:EulerEquations} correspond to Euler's equations of quantum hydrodynamics \cite{Petrov2003, Petrov2004} for the densitylike, $\rho(t,z)$, and velocitylike, $-\beta_{2}\,\partial\theta(t,z)/\partial t$, fields: Equation \eqref{Eq:EulerEquations-a} expresses the ``nonconservation'' of the ``current'' $\rho(t,z)\times-\beta_{2}\,\partial\theta(t,z)/\partial t$ under the effect of the losses---it is a \textit{bona fide} ``conservation'' equation only when $\alpha_{0}$ and $\alpha_{2}$ are zero---and the derivative of Equation~\eqref{Eq:EulerEquations-b} with respect to the spacelike coordinate $t$ corresponds to Newton's second law in an ``energy'' landscape given by the opposite of the right-hand side of Equation~\eqref{Eq:EulerEquations-b}, the first (second, third) term of which being the equivalent of the so-called quantum potential (the kinetic energy, the interaction energy).

\section{Monochromatic beam}
\label{Sec:MonochromaticBeam}

In this section, we solve Equations~\eqref{Eq:EulerEquations} in the ideal configuration where the beam of light is monochromatic at $\mathrm{\Omega}$. Accordingly, the slowly varying amplitude $A(t,z)$ of the complex electric field \eqref{Eq:ComplexElectricField} does not depend on time:
\begin{equation}
\label{Eq:MonochromaticLongitudinalEnvelope}
A(t,z)=A_{0}(z)=\sqrt{\rho_{0}(z)}\,e^{i\theta_{0}(z)},
\end{equation}
where $\rho_{0}(z)$ and $\theta_{0}(z)$ are solutions of the $t$-independent versions of Equations~\eqref{Eq:EulerEquations}, i.e.,
\begin{subequations}
\label{Eq:MonochromaticEulerEquations}
\begin{align}
\label{Eq:MonochromaticEulerEquations-a}
\frac{d\rho_{0}}{dz}&=-(\alpha_{0}+\alpha_{2}\,\rho_{0})\,\rho_{0}, \\
\label{Eq:MonochromaticEulerEquations-b}
\frac{d\theta_{0}}{dz}&=\gamma\,\rho_{0}.
\end{align}
\end{subequations}

\begin{figure}[t!]
\includegraphics[width=\linewidth]{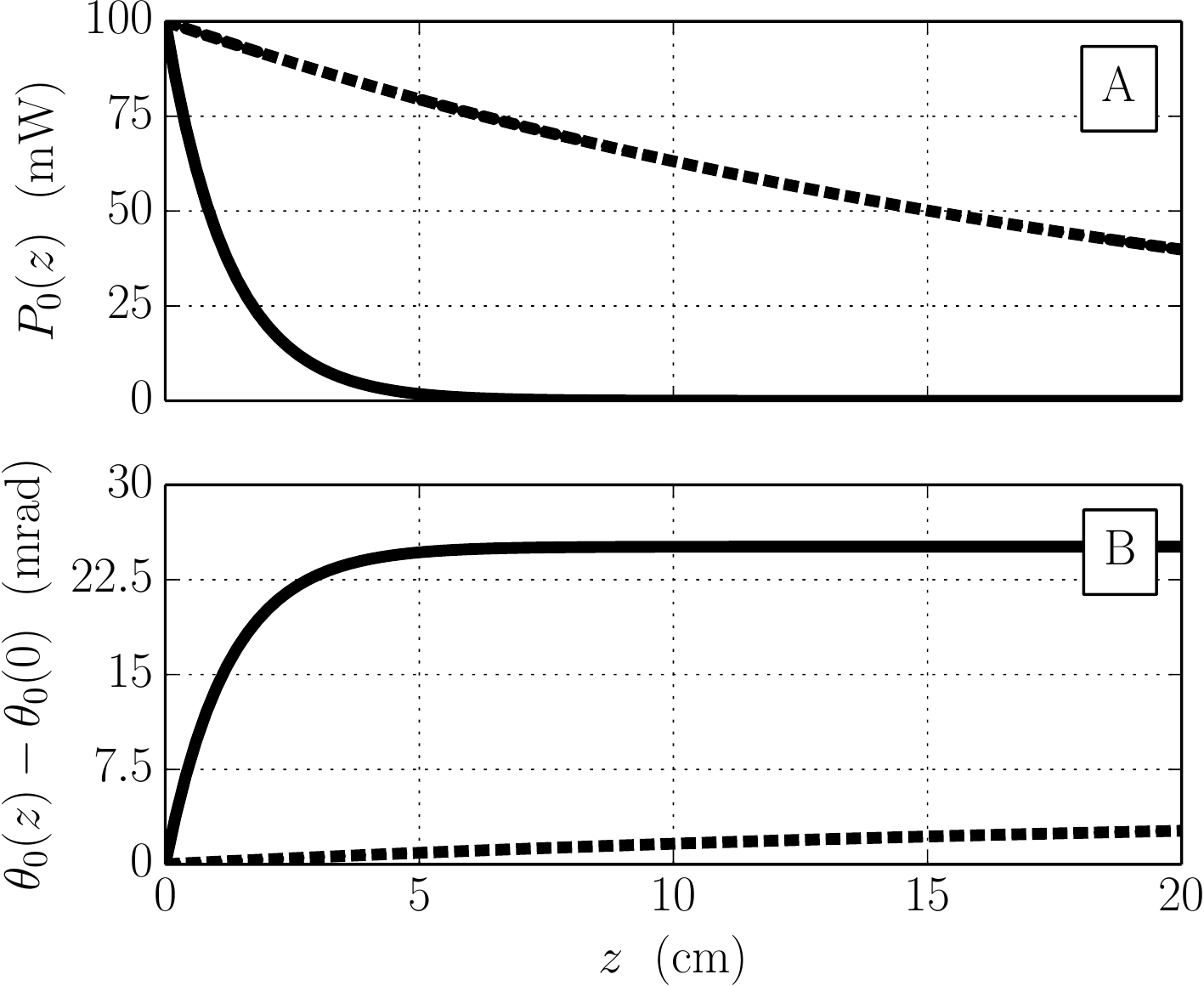}
\caption{Power $P_{0}(z)=\frac{1}{2}\,c\,\varepsilon_{0}\,n_{0}\,\rho_{0}(z)$ [Panel A; Equation~\eqref{Eq:MonochromaticDensity}] and phase $\theta_{0}(z)-\theta_{0}(0)$ [Panel B; Equations~\eqref{Eq:MonochromaticPhase}] of the beam of monochromatic light as a function of the propagation distance $z\in[0,20~\mathrm{cm}]$. The plain (dashed) curves are obtained for a TM mode at $1.55~\mu\mathrm{m}$ propagating along a channel waveguide with a silicon (silicon-nitride) core, the optical parameters of which are listed in the middle (right) column of Table~\ref{Tab:ExperimentalParameters}.}
\label{Fig:PowerPhase}
\end{figure}

\begin{table*}[t!]
\caption{Optical parameters at the $1.55~\mu\mathrm{m}$ telecommunication wavelength of single-mode channel waveguides whose cores are made of silicon (middle column) and of silicon nitride (right column). These figures are obtained from numerical integrations of the differential equations describing the nondegenerate four-wave mixing in silicon-based optical waveguides (see, e.g., Reference~\cite{Gao2011}). The simulations are based on the finite-element-method engine of COMSOL Multiphysics\textsuperscript{\scriptsize{\textregistered}}. In the table below, ``TM'' (``TE'') stands for ``transverse magnetic'' (``transverse electric''), indicating a configuration for which there is no magnetic (electric) field in the direction of propagation. Note that the negative (positive) $\beta_{2}$ in the TM-mode (TE-mode) silicon-nitride case is obtained for a striped (standard) configuration with air (silica) cladding.}
\label{Tab:ExperimentalParameters}
\begin{center}
\begin{tabular}{lll}
\hline\hline
\noalign{\smallskip}
& Silicon ($\text{Si}$) waveguide & Silicon-nitride ($\text{Si}_{\text{3}}\text{N}_{\text{4}}$) waveguide \\
\noalign{\smallskip}
\hline
\noalign{\smallskip}
Effective refractive index, $n_{0}$ & $\simeq\begin{cases}1.8\!&\!\text{(TM mode)}\\2.3\!&\!\text{(TE mode)}\end{cases}$ & $\simeq\begin{cases}1.5\!&\!\text{(TM mode)}\\1.6\!&\!\text{(TE mode)}\end{cases}$ \\
\noalign{\smallskip}
Group-velocity-dispersion parameter, $\beta_{2}$ & $\simeq\begin{cases}13.0~\mathrm{ps}^{2}\cdot\mathrm{m}^{-1}\!&\!\text{(TM mode)}\\-1.3~\mathrm{ps}^{2}\cdot\mathrm{m}^{-1}\!&\!\text{(TE mode)}\end{cases}$ & $\simeq\begin{cases}-0.6~\mathrm{ps}^{2}\cdot\mathrm{m}^{-1}\!&\!\text{(TM mode)}\\0.3~\mathrm{ps}^{2}\cdot\mathrm{m}^{-1}\!&\!\text{(TE mode)}\end{cases}$ \\
\noalign{\smallskip}
Kerr index, $n_{2}$ & $\sim\frac{1}{2}\,c\,\varepsilon_{0}\,n_{0}\times10^{-18}~\mathrm{m}^{2}\cdot\mathrm{V}^{-2}$ & $\sim\frac{1}{2}\,c\,\varepsilon_{0}\,n_{0}\times10^{-19}~\mathrm{m}^{2}\cdot\mathrm{V}^{-2}$ \\
\noalign{\smallskip}
Effective mode area, $A_{\mathrm{eff}}$ & $\simeq0.2~\mu\mathrm{m}^{2}$ & $\simeq2.0~\mu\mathrm{m}^{2}$ \\
\noalign{\smallskip}
Kerr-nonlinearity coefficient, $\gamma$ & $\simeq\frac{1}{2}\,c\,\varepsilon_{0}\,n_{0}\times20.3~\mathrm{m}^{-1}\cdot\mathrm{V}^{-2}$ & $\simeq\frac{1}{2}\,c\,\varepsilon_{0}\,n_{0}\times20.3\times10^{-2}~\mathrm{m}^{-1}\cdot\mathrm{V}^{-2}$ \\
\noalign{\smallskip}
One-photon-loss coefficient, $\alpha_{0}$ & $\simeq3.5~\mathrm{dB}\cdot\mathrm{cm}^{-1}$ & $\simeq0.2~\mathrm{dB}\cdot\mathrm{cm}^{-1}$ \\
\noalign{\smallskip}
Two-photon-loss coefficient, $\alpha_{2}$ & $\simeq0.2\,|\gamma|$ & $\longrightarrow0$ \\
\noalign{\smallskip}
\hline\hline
\end{tabular}
\end{center}
\end{table*}

Integrating Equations~\eqref{Eq:MonochromaticEulerEquations} yields the following analytical expressions for $\rho_{0}(z)$ and $\theta_{0}(z)$ as a function of the propagation coordinate $z\in[0,L]$ along the waveguide of length $L$:
\begin{equation}
\label{Eq:MonochromaticDensity}
\rho_{0}(z)=\rho_{0}(0)\,\frac{e^{-\alpha_{0}z}}{\displaystyle{1+\frac{\alpha_{2}\,\rho_{0}(0)}{\alpha_{0}}\,(1-e^{-\alpha_{0}z})}}
\end{equation}
and, defining the space average $\langle f(z)\rangle_{z}=\frac{1}{z}\int_{0}^{z}dz'\,f(z')$ of some $z$-dependent quantity $f(z)$,
\begin{subequations}
\label{Eq:MonochromaticPhase}
\begin{align}
\label{Eq:MonochromaticPhase-a}
\theta_{0}(z)&=\theta_{0}(0)+\gamma\,\langle\rho_{0}(z)\rangle_{z}\,z, \\
\label{Eq:MonochromaticPhase-b}
\langle\rho_{0}(z)\rangle_{z}&=\frac{1}{\alpha_{2}\,z}\ln\!\bigg[1+\frac{\alpha_{2}\,\rho_{0}(0)}{\alpha_{0}}\,(1-e^{-\alpha_{0}z})\bigg].
\end{align}
\end{subequations}
The graphical representations of Equations~\eqref{Eq:MonochromaticDensity} and \eqref{Eq:MonochromaticPhase} are shown in Figure~\ref{Fig:PowerPhase} for typical nonlinear-silicon-photonics \cite{Leuthold2010, Vivien2013} optical parameters in the $1.55~\mu\mathrm{m}$ telecommunication range. We particularly consider the propagation of a TM mode along waveguides with silicon (plain curves) and silicon-nitride (dashed curves) cores, the optical constants of them are listed in the middle and right columns of Table~\ref{Tab:ExperimentalParameters}. The silicon $\rho_{0}(z)$ and $\theta_{0}(z)$ vary more rapidly than the silicon-nitride ones because one-photon losses---that give $1/\alpha_{0}$ as one of the typical scales of variation for $\rho_{0}(z)$ and $\theta_{0}(z)$ [cf.~Equations~\eqref{Eq:MonochromaticDensity} and \eqref{Eq:MonochromaticPhase}]---is more important in silicon than in silicon nitride (see the next-to-the-last row of Table~\ref{Tab:ExperimentalParameters}).

\section{Bogoliubov dispersion relation}
\label{Sec:BogoliubovDispersionRelation}

In this section, we investigate the dispersion relation of weak-amplitude deviations from the $t$-independent and $z$-dependent power-phase pattern given in Equations~\eqref{Eq:MonochromaticDensity} and \eqref{Eq:MonochromaticPhase}. In Section~\ref{SubSec:DissipativeBogoliubovDeGennesEquations}, we derive the propagation equations of these fluctuations within Bogoliubov's theory of elementary excitations. We then solve them in Sections~\ref{SubSec:LosslessWaveguide} and \ref{SubSec:LossyWaveguide} and get the corresponding dispersion law---the so-called Bogoliubov dispersion relation---in, respectively, the lossless ($\alpha_{0},\alpha_{2}=0$) and the lossy ($\alpha_{0},\alpha_{2}\neq0$) configuration. For a CW beam of light, time translational symmetry makes the Bogoliubov dispersion relation be a function of the angular frequency $\omega$ of the modulation of the incident beam's complex amplitude, at $z=0$. The Bogoliubov law measured after propagation along the waveguide, at $z=L$, is the key quantity of the experiment proposed and detailed in Section~\ref{Sec:Experiment}.

\subsection{Dissipative Bogoliubov-de Gennes equations}
\label{SubSec:DissipativeBogoliubovDeGennesEquations}

Let us consider $t$-dependent departures from the steady profiles \eqref{Eq:MonochromaticDensity} and \eqref{Eq:MonochromaticPhase}. This amounts to search for the solutions $\rho(t,z)$ and $\theta(t,z)$ of Equations~\eqref{Eq:EulerEquations} in the form
\begin{align}
\label{Eq:DensityExpansion}
\rho(t,z)&=\rho_{0}(z)+\varrho(t,z), \\
\label{Eq:PhaseExpansion}
\theta(t,z)&=\theta_{0}(z)+\vartheta(t,z),
\end{align}
where $\varrho(t,z)$ and $-\beta_{2}\,\partial\vartheta(t,z)/\partial t$ are real fluctuating fields that are in addition assumed to be small \cite{Petrov2003, Petrov2004}. Inserting Equations~\eqref{Eq:DensityExpansion} and \eqref{Eq:PhaseExpansion} into Equations~\eqref{Eq:EulerEquations}, linearizing the corresponding system around $[\rho(t,z),-\beta_{2}\,\partial\theta(t,z)/\partial t]=[\rho_{0}(z),0]$, and Fourier expanding $\varrho(t,z)$ and $\vartheta(t,z)$ as \cite{Petrov2003, Petrov2004}
\begin{align}
\notag
\varrho(t,z)&\left.=\sqrt{\rho_{0}(z)}\int_{-\infty}^{\infty}\frac{d\omega}{2\pi}\,[a(\omega)\,f_{+}^{\vphantom{\ast}}(\omega,z)\,e^{-i\omega t}\right. \\
\label{Eq:DensityFluctuation}
&\left.\hphantom{=}+a^{\ast}(\omega)\,f_{+}^{\ast}(\omega,z)\,e^{i\omega t}],\right. \\
\notag
\vartheta(t,z)&\left.=\frac{1}{2\,i\,\sqrt{\rho_{0}(z)}}\int_{-\infty}^{\infty}\frac{d\omega}{2\pi}\,[a(\omega)\,f_{-}^{\vphantom{\ast}}(\omega,z)\,e^{-i\omega t}\right. \\
\label{Eq:PhaseFluctuation}
&\left.\hphantom{=}-a^{\ast}(\omega)\,f_{-}^{\ast}(\omega,z)\,e^{i\omega t}],\right.
\end{align}
we straightforwardly obtain the following matrix differential equation for the Fourier amplitudes $f_{\pm}(\omega,z)$:
\begin{subequations}
\label{Eq:BogoliubovDeGennesEquations}
\begin{align}
\label{Eq:BogoliubovDeGennesEquations-a}
i\,\frac{\partial}{\partial z}
\begin{bmatrix}
f_{+} \\ f_{-}
\end{bmatrix}
&=-\mathcal{K}
\begin{bmatrix}
f_{+} \\ f_{-}
\end{bmatrix}
, \\
\label{Eq:BogoliubovDeGennesEquations-b}
\mathcal{K}(\omega,z)&=
\begin{bmatrix}
\displaystyle{\frac{i}{2}\,[\alpha_{0}+3\,\alpha_{2}\,\rho_{0}(z)]} &
\displaystyle{\frac{\beta_{2}}{2}\,\omega^{2}} \vspace{1mm} \\
\displaystyle{\frac{\beta_{2}}{2}\,\omega^{2}+2\,\gamma\,\rho_{0}(z)} &
\displaystyle{\frac{i}{2}\,[\alpha_{0}+\alpha_{2}\,\rho_{0}(z)]}
\end{bmatrix}
.
\end{align}
\end{subequations}

The $a(\omega)$'s in Equations~\eqref{Eq:DensityFluctuation} and \eqref{Eq:PhaseFluctuation} are chosen to be $z$ independent and homogeneous to a voltage times a time so that the $f_{\pm}(\omega,z)$'s encapsulate all the $z$ dependence of the fluctuations and are by construction dimensionless. In the absence of photon losses ($\alpha_{0},\alpha_{2}=0$), Equation~\eqref{Eq:BogoliubovDeGennesEquations-a} and the opposite of Equation~\eqref{Eq:BogoliubovDeGennesEquations-b} are formally analogous to the Bogoliubov-de Gennes matrix equation and the Bogoliubov-de Gennes Hamiltonian of dilute atomic Bose gases within Madelung's picture \cite{Petrov2003, Petrov2004}. When present, one-photon losses ($\alpha_{0}$ terms) enter the matrix propagation equation \eqref{Eq:BogoliubovDeGennesEquations-a} of the fluctuations into diagonal terms, in a similar fashion to the gain of a laser medium in the matrix propagation equation of the modulations of a paraxial optical field \cite{Boyd1992}; while accounting for the effect of a gain would indeed correspond to consider a diagonal propagation term of the form $(\partial f_{\pm}/\partial z)_{\mathrm{Gain}}=G\,f_{\pm}$, with $G>0$, thus describing an amplification of the modulations of the optical field, the diagonal one-photon-loss terms come out with a ``$-$'' sign in Equation~\eqref{Eq:BogoliubovDeGennesEquations-a} and have on the contrary the tendency to make the amplitude of the modulations decrease in the course of the propagation of the beam of light along the waveguide. On the other hand, two-photon losses ($\alpha_{2}$ terms) enter Equation~\eqref{Eq:BogoliubovDeGennesEquations-a} into intensity-dependent diagonal terms exactly acting as gain-saturation terms in laser media. Finally, note that as $\mathcal{K}(\omega,z)$ is an even function of $\omega$, its eigenelements determined after will be so too.

\subsection{Lossless waveguide}
\label{SubSec:LosslessWaveguide}

Let us first analyze the simplest situation where $\alpha_{0},\alpha_{2}=0$. In this ideal, lossless, case, there is no power loss in the course of the propagation of the beam of light along the waveguide:
\begin{equation}
\label{Eq:TrivialDensity}
\rho_{0}(z)\overset{\eqref{Eq:MonochromaticDensity}}{=}\rho_{0}(0)=\rho_{0}=\mathrm{const},
\end{equation}
and the phase of the beam consequently grows linearly with the propagation distance $z$:
\begin{equation}
\label{Eq:TrivialPhase}
\theta_{0}(z)\overset{\eqref{Eq:MonochromaticPhase}}{=}\theta_{0}(0)+\gamma\,\rho_{0}\,z.
\end{equation}

Accordingly, in addition to having zero diagonal terms, the Bogoliubov-de Gennes-type matrix \eqref{Eq:BogoliubovDeGennesEquations-b} is homogeneous, so that the Fourier components $f_{\pm}(\omega,z)$ of the density and phase fluctuations of the fluid of light, solutions of Equation~\eqref{Eq:BogoliubovDeGennesEquations-a}, are plane waves with $\omega$-dependent amplitudes $\tilde{f}_{\pm}(\omega)$ and wave number $k(\omega)$ along the propagation, $z$, axis:
\begin{equation}
\label{Eq:PlaneWaveSolution}
f_{\pm}(\omega,z)=\tilde{f}_{\pm}(\omega)\,e^{ik(\omega)z}.
\end{equation}
Inserting the physical ansatz \eqref{Eq:PlaneWaveSolution} into the differential equation \eqref{Eq:BogoliubovDeGennesEquations-a} straightforwardly yields the eigenelement problem
\begin{subequations}
\label{Eq:EigenvalueProblem}
\begin{align}
\label{Eq:EigenvalueProblem-a}
k(\omega)
\begin{bmatrix}
\tilde{f}_{+}(\omega) \\ \tilde{f}_{-}(\omega)
\end{bmatrix}
&=
\mathcal{K}(\omega)
\begin{bmatrix}
\tilde{f}_{+}(\omega) \\ \tilde{f}_{-}(\omega)
\end{bmatrix}
, \\
\label{Eq:EigenvalueProblem-b}
\mathcal{K}(\omega)&=
\begin{bmatrix}
0 & \displaystyle{\frac{\beta_{2}}{2}\,\omega^{2}} \vspace{1mm} \\
\displaystyle{\frac{\beta_{2}}{2}\,\omega^{2}+2\,\gamma\,\rho_{0}} & 0
\end{bmatrix}
.
\end{align}
\end{subequations}
Its two eigenvalues $k(\omega)$, roots of the characteristic polynomial $\det[\mathcal{K}(\omega)-X\,\mathds{1}_{2}]=X^{2}+\det[\mathcal{K}(\omega)]$, are by construction symmetrically opposite---corresponding to a positive, ``$+$,'' branch and a negative, ``$-$,'' one \cite{Castin2001}---and read
\begin{subequations}
\label{Eq:PlaneWaveWavenumber}
\begin{align}
\label{Eq:PlaneWaveWavenumber-a}
k(\omega)&=\pm\,\sqrt{-\det[\mathcal{K}(\omega)]} \\
\label{Eq:PlaneWaveWavenumber-b}
&=\pm\,\sqrt{\frac{\beta_{2}}{2}\,\omega^{2}\,\bigg(\frac{\beta_{2}}{2}\,\omega^{2}+2\,\gamma\,\rho_{0}\bigg)}.
\end{align}
\end{subequations}
Equation \eqref{Eq:PlaneWaveWavenumber-b} is nothing but the wave-number--angular-frequency relation of the power and phase fluctuations, that is, by definition, the Bogoliubov dispersion relation, of the homogeneous beam of monochromatic light \eqref{Eq:TrivialDensity}, \eqref{Eq:TrivialPhase}. Note that this quantity may \textit{a priori} be complex, depending on the sign of the group-velocity-dispersion parameter $\beta_{2}$ and on the one of the Kerr-nonlinearity coefficient $\gamma$.

\begin{figure}[t!]
\includegraphics[width=\linewidth]{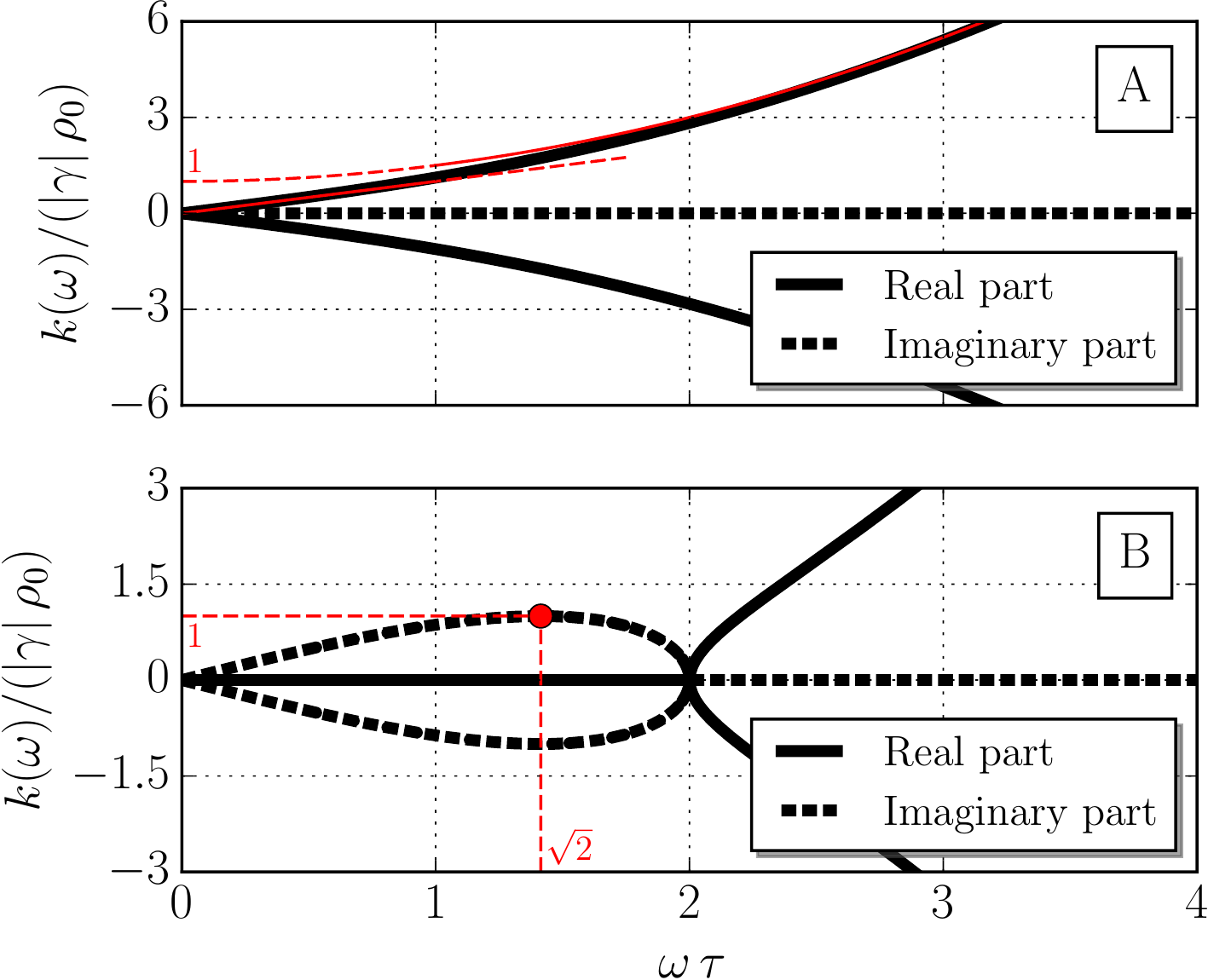}
\caption{Real (plain curves) and imaginary  (dashed curves) parts of the normalized Bogoliubov dispersion relation $k(\omega)/(|\gamma|\,\rho_{0})$ against $\omega\,\tau\geqslant0$ in the absence of one- and two-photon losses, as given by Equation~\eqref{Eq:PlaneWaveWavenumber-b}. The plots are symmetric with respect to the horizontal $k(\omega)=0$ line: The branches above (below) this line correspond to the ``$+$'' (``$-$'') sign in Equation~\eqref{Eq:PlaneWaveWavenumber-b} and are called ``positive (negative) branches.'' Panel A: ``Dynamically'' stable case [in opposition to the situation discussed after Equations~\eqref{Eq:PlaneWaveWavenumberCases}] where the group-velocity-dispersion parameter $\beta_{2}$ and the Kerr-nonlinearity coefficient $\gamma$ have the same sign; the lower (upper) red curve indicates the low-$\omega$ (large-$\omega$) linear (parabolic) behavior of the positive branch of the Bogoliubov dispersion relation, as determined in the first (second) row of Equation~\eqref{Eq:PlaneWaveWavenumberLimits}. Panel B: ``Dynamically'' unstable case [discussed after Equations~\eqref{Eq:PlaneWaveWavenumberCases}] where $\beta_{2}$ and $\gamma$ have opposite signs; the red marker at $(\sqrt{2},1)$ refers to the remarkable identity \eqref{Eq:PlaneWaveWavenumberCases-c} for the positive branch of the imaginary part of $k(\omega)/(|\gamma|\,\rho_{0})$.}
\label{Fig:DissipationlessSituation}
\end{figure}

When $\beta_{2}$ and $\gamma$ are of same sign, i.e., when $\beta_{2}>0$ (normal group-velocity dispersion) and $\gamma>0$ (self-focusing Kerr nonlinearity) or when $\beta_{2}<0$ (anomalous group-velocity dispersion) and $\gamma<0$ (self-defocusing Kerr nonlinearity), the dispersion law $k(\omega)$ given in Equation~\eqref{Eq:PlaneWaveWavenumber-b} is a real function of $\omega$ that directly corresponds, within the $t\longleftrightarrow z$ mapping (then within the $\text{angular-frequency}\longleftrightarrow\text{wave-number}$ mapping), to the dispersion relation of the elementary excitations propagating on top of a homogeneous dilute atomic Bose-Einstein condensate at rest \cite{Vocke2015, Larre2015b, Larre2016a}. The latter is linear, or ``phononlike'' \cite{Castin2001, Pethick2002, Pitaevskii2016} within the atomic-gas framework, at small $\omega$'s and quadratic, ``particlelike'' \cite{Castin2001, Pethick2002, Pitaevskii2016}, at large $\omega$'s:
\begin{equation}
\label{Eq:PlaneWaveWavenumberLimits}
k(\omega)\simeq\pm
\begin{cases}
v^{-1}\,|\omega|, & |\omega|\ll1/\tau, \\
\displaystyle{\frac{|\beta_{2}|}{2}\,\omega^{2}+|\gamma|\,\rho_{0}}, & |\omega|\gg1/\tau,
\end{cases}
\end{equation}
where the parameters
\begin{align}
\label{Eq:SpeedOfSound}
v^{-1}&=\sqrt{\beta_{2}\,\gamma\,\rho_{0}}, \\
\label{Eq:HealingLength}
\tau&=\sqrt{\frac{\beta_{2}}{\gamma\,\rho_{0}}}=\frac{|\beta_{2}|}{v^{-1}},
\end{align}
respectively homogeneous to the inverse of a velocity and to a time, are within the $t\longleftrightarrow z$ mapping the respective optical analogs of the Bogoliubov speed of sound and of the healing length \cite{Castin2001, Pethick2002, Pitaevskii2016} of a homogeneous dilute atomic Bose gas (recalling that $-1/\beta_{2}$, $-\gamma$, and $\rho_{0}$ play the role of the mass, of the interaction constant, and of the uniform density of the quantum fluid, respectively). We plot in Figure~\ref{Fig:DissipationlessSituation}A $k(\omega)/(|\gamma|\,\rho_{0})$ against $\omega\,\tau\geqslant0$, as given by Equation~\eqref{Eq:PlaneWaveWavenumber-b} for $\beta_{2}$'s and $\gamma$'s having the same sign. The red curves on this graph indicate the low- and large-$\omega$ approximations \eqref{Eq:PlaneWaveWavenumberLimits} for the positive, ``$+$,'' branch of the Bogoliubov dispersion relation. According to Landau's criterion for superfluidity \cite{Landau1941, Landau1980}, a zero-temperature conservative Bose-Einstein condensate flowing with a velocity smaller than the Bogoliubov speed of sound is energetically stable against the presence of a weakly perturbing impurity \cite{Pethick2002, Pitaevskii2016}; thus,
\begin{equation}
\label{Eq:CriticalVelocitySuperfluidity}
v^{-1}=\underset{\omega\in\mathbb{R}}{\min}\,\bigg|\frac{k(\omega)}{\omega}\bigg|
\end{equation}
defined in Equation~\eqref{Eq:SpeedOfSound} may be regarded as a direct optical equivalent of Landau's critical velocity for superfluidity, as recently investigated in the different optical configuration where a paraxial beam of monochromatic light propagates in a waveguide-free nonlinear medium \cite{Vocke2015}. Finally, note that the $\omega$-independent Hartree-type shift $\pm\,|\gamma|\,\rho_{0}$ in the second row of Equation~\eqref{Eq:PlaneWaveWavenumberLimits} corresponds to the naive nonlinearity-induced correction to the dispersion-induced fluctuation $\pm\,|\beta_{2}|\,\omega^{2}/2$ of the beam's propagation constant $\beta_{0}$; the rigorous Bogoliubov analysis carried out here shows that this is valid in the $|\omega|\gg1/\tau$ limit only and that the full propagation constant's fluctuation $k(\omega)$ is in fact gapless, i.e., that it vanishes at $\omega=0$, as one may verify in the first row of Equation~\eqref{Eq:PlaneWaveWavenumberLimits} and in Figure~\ref{Fig:DissipationlessSituation}A.

When $\beta_{2}$ and $\gamma$ are of opposite signs instead, $k(\omega)$ given in Equation~\eqref{Eq:PlaneWaveWavenumber-b} is a complex function of $\omega$ that we plot in Figure~\ref{Fig:DissipationlessSituation}B within the dimensionless units of Figure~\ref{Fig:DissipationlessSituation}A. In this case, one has the following noticeable identities:
\begin{subequations}
\label{Eq:PlaneWaveWavenumberCases}
\begin{align}
\label{Eq:PlaneWaveWavenumberCases-a}
\mathrm{Re}[k(\omega)]&=0\quad\text{for}\quad|\omega|<2/\tau, \\
\label{Eq:PlaneWaveWavenumberCases-b}
\mathrm{Im}[k(\omega)]&=0\quad\text{for}\quad|\omega|>2/\tau, \\
\label{Eq:PlaneWaveWavenumberCases-c}
|\mathrm{Im}[k(\pm\,\sqrt{2}/\tau)]|&=|\gamma|\,\rho_{0},
\end{align}
\end{subequations}
the latter being indicated  by means of a red marker for the positive branch of the Bogoliubov law's imaginary part. Within the quantum-fluid, ``Gross-Pitaevskii,'' framework, the fact that the Bogoliubov dispersion relation possesses an imaginary part signals that the evolving fluid of light is dynamically unstable \cite{Castin2001, Pethick2002, Pitaevskii2016}, very especially at the angular frequencies $\omega$ such that $\mathrm{Im}[k(\omega)]<0$, for which $|f_{\pm}(\omega,z)|=|\tilde{f}_{\pm}(\omega)|\,e^{-\mathrm{Im}[k(\omega)]z}$ diverges as the timelike parameter $z$ increases. Within the original nonlinear-optics, ``nonlinear Schr\"odinger,'' framework, this corresponds to the situation where the propagation of the light beam in the positive-$z$ direction is not robust against the formation of modulation instabilities (also called sideband instabilities) \cite{Boyd1992, Agrawal1995, Rosanov2002, Zakharov2009, Turitsyn2010, Conforti2014}. In that case, deviations from the background pattern \eqref{Eq:TrivialDensity}, \eqref{Eq:TrivialPhase} are reinforced by the Kerr optical nonlinearity of the underlying medium, leading to the generation of spectral sidebands and the eventual breakup of the wave profile into a train of pulses.

\subsection{Lossy waveguide}
\label{SubSec:LossyWaveguide}

Now, let us analyze the realistic configuration for which one- and two-photon losses occur at the operating angular frequency $\mathrm{\Omega}$: $\alpha_{0},\alpha_{2}\neq0$. In Section~\ref{SubSubSec:AdiabaticEvolution} first, we analytically investigate the case where the effective evolution of the Bogoliubov fluctuations in the positive-$z$ direction is adiabatic. To do so, we will make use of an optical version of the adiabatic theorem of quantum mechanics \cite{Born1928, Messiah1999, Sun1993}, derived in detail in Appendix \ref{App:AdiabaticTheoremForZDependentPropagatingOpticalSystems}. In Section~\ref{SubSubSec:ArbitraryEvolution} then, we numerically treat the general case where this effective evolution might possibly be nonadiabatic. We illustrate and discuss our results on the basis of the two concrete nonlinear-silicon-photonics examples of Table~\ref{Tab:ExperimentalParameters}.

\subsubsection{Adiabatic evolution}
\label{SubSubSec:AdiabaticEvolution}

A $z$-dependent configuration for which analytical solutions of the dissipative Bogoliubov-de Gennes differential system \eqref{Eq:BogoliubovDeGennesEquations-a} may still be easily obtained gets along with the case where the corresponding effective evolution along the propagation, $z$, axis is adiabatic. From Appendix \ref{App:AdiabaticTheoremForZDependentPropagatingOpticalSystems} and as mathematically formulated in the third paragraph of the present section, the constraint for having such an adiabatic effective evolution is that the nondiagonal elements of the rate of change of \eqref{Eq:BogoliubovDeGennesEquations-b} in the normalized basis of the \eqref{Eq:BogoliubovDeGennesEquations-b}'s eigenvectors and in units of the difference of the two \eqref{Eq:BogoliubovDeGennesEquations-b}'s eigenvalues must be smaller than this difference. In this case, each \eqref{Eq:BogoliubovDeGennesEquations-b}'s eigenvector is a local function of $z$ that strictly ``follows'' the variations of its corresponding eigenvalue as a function of $z$.

Accordingly, as the effective evolution \eqref{Eq:BogoliubovDeGennesEquations} is not cyclic, i.e., as $\mathcal{K}(\omega,L)\neq\mathcal{K}(\omega,0)$ [simply because $\rho_{0}(L)\neq\rho_{0}(0)$; see Equation~\eqref{Eq:MonochromaticDensity} or Figure~\ref{Fig:PowerPhase}A], Appendix \ref{App:AdiabaticTheoremForZDependentPropagatingOpticalSystems} demonstrates that the adiabatic solutions $f_{\pm}(\omega,z)$ of Equation~\eqref{Eq:BogoliubovDeGennesEquations-a} may be written in the generic form
\begin{subequations}
\label{Eq:NearlyPlaneWaveSolution}
\begin{align}
\label{Eq:NearlyPlaneWaveSolution-a}
f_{\pm}(\omega,z)&=\tilde{f}_{\pm}(\omega,z)\,e^{i\int_{0}^{z}dz'\,k(\omega,z')} \\
\label{Eq:NearlyPlaneWaveSolution-b}
&=\tilde{f}_{\pm}(\omega,z)\,e^{i\langle k(\omega,z)\rangle_{z}z},
\end{align}
\end{subequations}
where the local amplitudes $\tilde{f}_{\pm}(\omega,z)$ and the local wave number $k(\omega,z)$ along the $z$ axis are eigenelements of the two-by-two matrix $\mathcal{K}(\omega,z)$ given in Equation~\eqref{Eq:BogoliubovDeGennesEquations-b}:
\begin{equation}
\label{Eq:LocalEigenvalueProblem}
k(\omega,z)
\begin{bmatrix}
\tilde{f}_{+}(\omega,z) \\ \tilde{f}_{-}(\omega,z)
\end{bmatrix}
=
\mathcal{K}(\omega,z)
\begin{bmatrix}
\tilde{f}_{+}(\omega,z) \\ \tilde{f}_{-}(\omega,z)
\end{bmatrix}
.
\end{equation}
Equation \eqref{Eq:LocalEigenvalueProblem} admits nontrivial solutions when $k(\omega,z)$ is a root of the characteristic polynomial $\det[\mathcal{K}(\omega,z)-X\,\mathds{1}_{2}]=X^{2}-\mathrm{tr}[\mathcal{K}(\omega,z)]\,X+\det[\mathcal{K}(\omega,z)]$, i.e., when
\begin{subequations}
\label{Eq:CharacteristicEquation}
\begin{align}
\label{Eq:CharacteristicEquation-a}
k(\omega,z)&\left.=\frac{\mathrm{tr}[\mathcal{K}(\omega,z)]}{2}\pm\sqrt{\frac{\mathrm{tr}^{2}[\mathcal{K}(\omega,z)]}{4}-\det[\mathcal{K}(\omega,z)]}\right. \\
\notag
&\left.=i\,\bigg[\frac{\alpha_{0}}{2}+\alpha_{2}\,\rho_{0}(z)\bigg]\right. \\
\label{Eq:CharacteristicEquation-b}
&\left.\hphantom{=}\pm\sqrt{\frac{\beta_{2}}{2}\,\omega^{2}\,\bigg[\frac{\beta_{2}}{2}\,\omega^{2}+2\,\gamma\,\rho_{0}(z)\bigg]-\frac{[\alpha_{2}\,\rho_{0}(z)]^{2}}{4}},\right.
\end{align}
\end{subequations}
from which we deduce the Bogoliubov dispersion relation $\langle k(\omega,z)\rangle_{z}$ of the adiabatically evolving fluid of light, as appearing through Equation~\eqref{Eq:NearlyPlaneWaveSolution-b}.

These results hold when the adiabatic constraint textually formulated in the first paragraph of the present section is satisfied. By analogy with Equation~\eqref{Eq:ExactAdiabaticCondition}, this condition may be written in the form
\begin{align}
\notag
&\left.\underset{z\in[0,L]}{\max}\,\frac{\displaystyle{\bigg|\langle\tilde{f}_{\pm}(\omega,z)|\,\frac{\partial\mathcal{K}}{\partial z}(\omega,z)\,|\tilde{f}_{\mp}(\omega,z)\rangle\bigg|}}{|k_{\mp}(\omega,z)-k_{\pm}(\omega,z)|}\right. \\
\label{Eq:ExactAdiabaticConditionBdG}
&\left.\quad\ll\underset{z\in[0,L]}{\min}\,|k_{\mp}(\omega,z)-k_{\pm}(\omega,z)|,\right.
\end{align}
where $k_{\pm}(\omega,z)$ refers to the ``$\pm$'' branch of $k(\omega,z)$ in Equation~\eqref{Eq:CharacteristicEquation-b} and $|\tilde{f}_{\pm}(\omega,z)\rangle\propto{^{t}[\tilde{f}_{+,\pm}(\omega,z)~\tilde{f}_{-,\pm}(\omega,z)]}$ to the corresponding eigenvector, normalized to unity. In the very particular case where one- and two-photon losses are absent, i.e., in the case where $\alpha_{0},\alpha_{2}=0$, $\partial\mathcal{K}(\omega,z)/\partial z$ identically vanishes, making the left-hand side of \eqref{Eq:ExactAdiabaticConditionBdG} zero, and then the latter inequality perfectly satisfied, as it has to be in such a configuration; accordingly, the Bogoliubov dispersion relation $\langle k(\omega,z)\rangle_{z}$ reduces to $k(\omega)$ given in Equation~\eqref{Eq:PlaneWaveWavenumber-b}, as one readily checks from Equation~\eqref{Eq:CharacteristicEquation-b}.

\subsubsection{Arbitrary evolution}
\label{SubSubSec:ArbitraryEvolution}

When the effective evolution of the Bogoliubov fluctuations in the positive-$z$ direction is not adiabatic, the results derived in Section~\ref{SubSubSec:AdiabaticEvolution} do not hold, as a consequence of which one generically has to rely on a numerical resolution of the dissipative Bogoliubov-de Gennes-type problem \eqref{Eq:BogoliubovDeGennesEquations}. This is what we do in the next paragraph.

\begin{figure*}[t!]
\includegraphics[width=\linewidth]{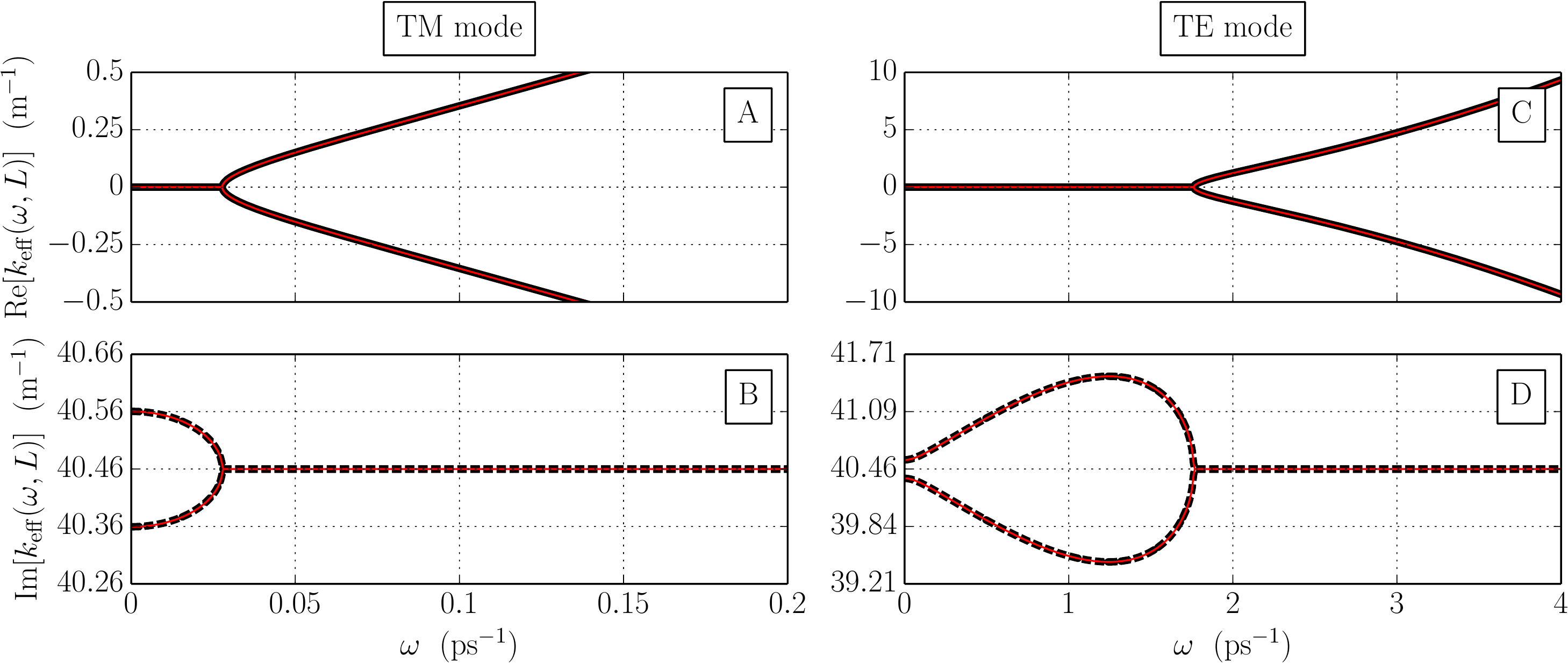}
\caption{Real (upper row; black plain style as in Figure~\ref{Fig:DissipationlessSituation}) and imaginary (lower row; black dashed style as in Figure~\ref{Fig:DissipationlessSituation}) parts of the Bogoliubov dispersion relation $k_{\mathrm{eff}}(\omega\geqslant0,z=L)$ of ``TM'' (left column) and ``TE'' (right column) fluids of light exiting a $L=2~\mathrm{cm}$-long silicon-core single-mode channel waveguide. The plots result from the numerical diagonalization of $\mathcal{K}_{\mathrm{eff}}(\omega,L)$ defined in Equation~\eqref{Eq:EffectiveBdGHamiltonian-b} and the red curves indicate the adiabatic predictions of Section~\ref{SubSubSec:AdiabaticEvolution}. The operating wavelength equals $1.55~\mu\mathrm{m}$, the incident power is of $100~\mathrm{mW}$, and the corresponding silicon's parameters are given in the middle column of Table~\ref{Tab:ExperimentalParameters}. The dispersions are horizontally symmetric: The upper (lower) branches correspond to the ``$+$'' (``$-$'') sign in the second row of Equation~\eqref{Eq:CharacteristicEquation-b} and are called in the text ``positive (negative) branches.''}
\label{Fig:DissipativeSituationSi}
\end{figure*}

\begin{figure*}[t!]
\includegraphics[width=\linewidth]{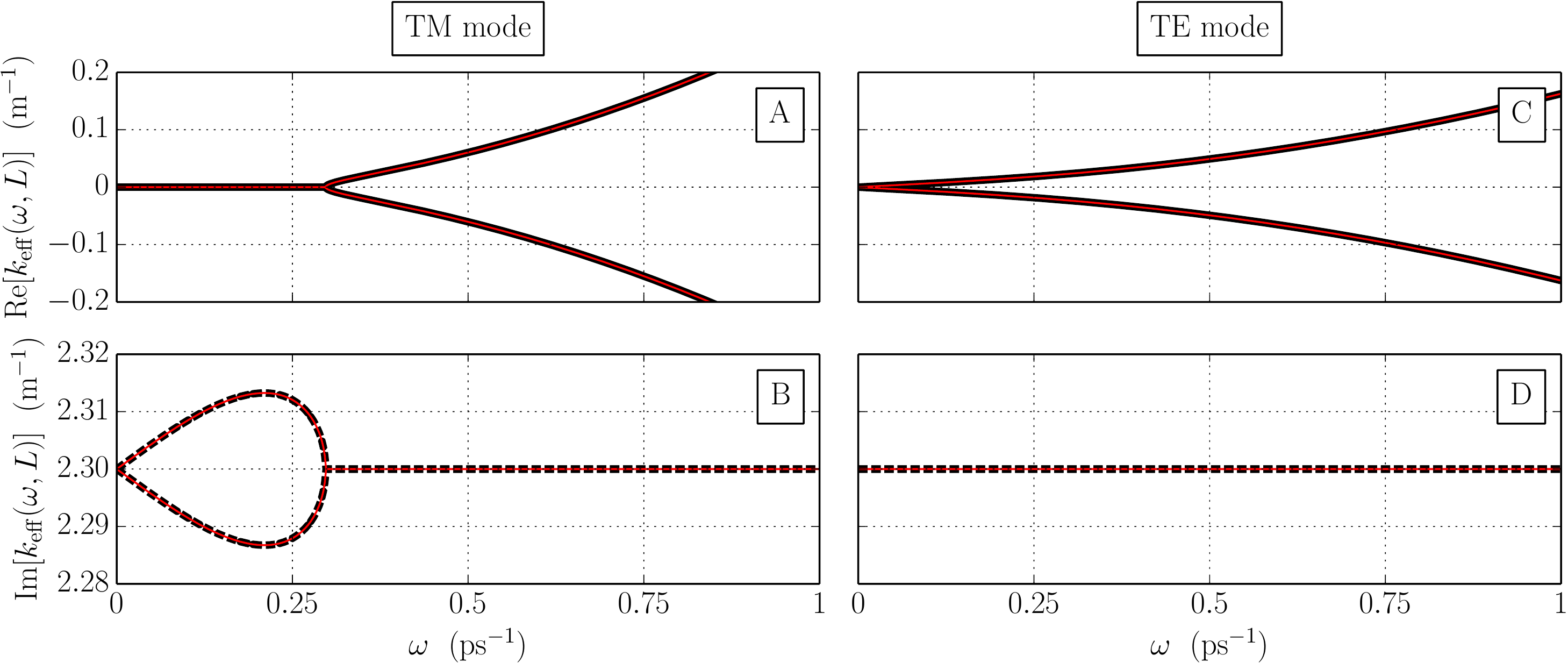}
\caption{Same as Figure~\ref{Fig:DissipativeSituationSi} for a $L=20~\mathrm{cm}$-long silicon-nitride-core single-mode channel waveguide, the parameters of which are given in the right column of Table~\ref{Tab:ExperimentalParameters}.}
\label{Fig:DissipativeSituationSiNi}
\end{figure*}

We start by writing the general solution of Equation~\eqref{Eq:BogoliubovDeGennesEquations-a} in the formal matrix exponential form
\begin{equation}
\label{Eq:FormalMatrixExponentialForm}
\begin{bmatrix}
f_{+}(\omega,z) \\ f_{-}(\omega,z)
\end{bmatrix}
=\exp[i\,\mathcal{K}_{\mathrm{eff}}(\omega,z)\,z]
\begin{bmatrix}
f_{+}(\omega,0) \\ f_{-}(\omega,0)
\end{bmatrix}
,
\end{equation}
where the $\omega$-, $z$-dependent two-by-two matrix $\mathcal{K}_{\mathrm{eff}}(\omega,z)$ is defined through
\begin{subequations}
\label{Eq:EffectiveBdGHamiltonian}
\begin{align}
\label{Eq:EffectiveBdGHamiltonian-a}
\exp[i\,\mathcal{K}_{\mathrm{eff}}(\omega,z)\,z]&=Z\bigg\{\!\exp\!\bigg[i\int_{0}^{z}dz'\,\mathcal{K}(\omega,z')\bigg]\bigg\} \\
\label{Eq:EffectiveBdGHamiltonian-b}
&=\underset{N\to\infty}{\lim}\prod_{n=N}^{0}\exp[i\,\mathcal{K}(\omega,z_{n})\,\delta z].
\end{align}
\end{subequations}
In Equation~\eqref{Eq:EffectiveBdGHamiltonian-a}, $Z\{\cdot\}$ is the equivalent of the chronological ordering \cite{Weinberg1995} for time-dependent quantum-mechanical systems; it standardly appears because $\mathcal{K}(\omega,z)\,\mathcal{K}(\omega,z')\neq\mathcal{K}(\omega,z')\,\mathcal{K}(\omega,z)$ for all $z'\neq z$ and may be defined through the infinite, reversely ordered product \eqref{Eq:EffectiveBdGHamiltonian-b}, where $z_{n}=n\,\delta z=n\,z/N$ ($n\in\llbracket0,N\rrbracket$). In this case, the Bogoliubov dispersion relation of the fluid of light corresponds to the local eigenvalues $k_{\mathrm{eff}}(\omega,z)$ of the effective propagation matrix $\mathcal{K}_{\mathrm{eff}}(\omega,z)$, that we determine from the numerical diagonalization of the latter by means of Equation~\eqref{Eq:EffectiveBdGHamiltonian-b}. We plot in Figures~\ref{Fig:DissipativeSituationSi} and \ref{Fig:DissipativeSituationSiNi} the real (upper panels; black plain style as in Figure~\ref{Fig:DissipationlessSituation}) and imaginary (lower panels; black dashed style as in Figure~\ref{Fig:DissipationlessSituation}) parts of $k_{\mathrm{eff}}(\omega\geqslant0,z=L)$ in the case of a $L=2~\mathrm{cm}$-long silicon-core and a $L=20~\mathrm{cm}$-long silicon-nitride-core, respectively, single-mode channel waveguide supporting a TM (left panels) or a TE (right panels) mode. The incident light beam operates at $1.55~\mu\mathrm{m}$ and $100~\mathrm{mW}$, and the waveguides' parameters are given in Table~\ref{Tab:ExperimentalParameters}. As stated in the introduction of Section~\ref{Sec:BogoliubovDispersionRelation}, the Bogoliubov dispersion relation $k_{\mathrm{eff}}(\omega,z)$ is here displayed at $z=L$ because the beam of light is typically imaged at the exit of the waveguide, precisely where $z=L$.

By plotting in red the real and imaginary parts of the adiabatic Bogoliubov dispersion relation $\langle k(\omega,z)\rangle_{L}$ on top of the exact $\mathrm{Re}[k_{\mathrm{eff}}(\omega,L)]$ and $\mathrm{Im}[k_{\mathrm{eff}}(\omega,L)]$, we note that the agreement between the predictions of Section~\ref{SubSubSec:AdiabaticEvolution} and the present numerical results is good, although the two nonlinear-silicon-photonics examples examinated here fall into the \textit{a priori} unfavorable situation where one-photon losses dominate the Kerr nonlinearity: For the waveguide made of a silicon (silicon-nitride) core, $\gamma\,\rho_{0}(0)\simeq2.0~\mathrm{m}^{-1}$ ($\simeq2.0\times10^{-2}~\mathrm{m}^{-1}$) while $\alpha_{0}/2\simeq40.3~\mathrm{m}^{-1}$ ($\simeq2.3~\mathrm{m}^{-1}$). We numerically check from Equations~\eqref{Eq:BogoliubovDeGennesEquations-b} and \eqref{Eq:LocalEigenvalueProblem} that the adiabatic-evolution constraint \eqref{Eq:ExactAdiabaticConditionBdG} is verified for a wide range of Bogoliubov angular frequencies $\omega$ and that the ratio of its left-hand side by its right-hand one is as small as $1/\omega^{4}$ is when $\omega$ is large. Accordingly, in the discussions below, we will use the adiabatic identification
\begin{equation}
\label{Eq:AdiabaticApproximation}
k_{\mathrm{eff}}(\omega,L)=\langle k(\omega,z)\rangle_{L}=\frac{1}{L}\int_{0}^{L}dz\;k(\omega,z)
\end{equation}
to quantitatively describe the silicon and silicon-nitride Bogoliubov dispersions shown in Figures~\ref{Fig:DissipativeSituationSi} and \ref{Fig:DissipativeSituationSiNi}.

Figures~\ref{Fig:DissipativeSituationSi}A, \ref{Fig:DissipativeSituationSi}C, and \ref{Fig:DissipativeSituationSiNi}A (\ref{Fig:DissipativeSituationSi}B, \ref{Fig:DissipativeSituationSi}D, and \ref{Fig:DissipativeSituationSiNi}B) display the same qualitative behavior for the real (imaginary) part of the Bogoliubov dispersion relation of the fluid of light exiting the waveguide. In sharp contrast to the linear dispersion of the propagating phononlike mode in the lossless ($\alpha_{0}$ and $\alpha_{2}$ null), stable ($\beta_{2}$ and $\gamma$ of same sign) configuration of Section~\ref{SubSec:LosslessWaveguide} [see the first row of Equation~\eqref{Eq:PlaneWaveWavenumberLimits} and Figure~\ref{Fig:DissipationlessSituation}A], the Bogoliubov fluctuations show here an overdamped, nonpropagating, behavior at low $\omega$, as already described theoretically in the context of semiconductor-microcavity exciton-polariton quantum fluids \cite{Wouters2007, Wouters2008, Wouters2009, Byrnes2012, Larre2012, Carusotto2013, Larre2013}. In this regime indeed, the real part of the Bogoliubov law is dispersionless, equal to zero. The latter starts being nonzero and symmetric with respect to zero at large $\omega$ while the imaginary part becomes $\omega$ independent. The latter remains anyhow nonzero and positive, as it is in the low-$\omega$ regime where it is symmetric with respect to its large-$\omega$ value. This positiveness indicates that the Bogoliubov waves oscillating on top of the fluid of light are exponentially damped at any $\omega$ according to $e^{-\mathrm{Im}[k_{\mathrm{eff}}(\omega,L)]L}$. On the other hand, the noticed horizontal symmetry of the curves directly refers to the ``$\pm$'' sign in the second row of Equation~\eqref{Eq:CharacteristicEquation-b}: For each plot, the upper, ``$+$,'' branch corresponds to what we call the positive branch of the Bogoliubov dispersion relation and the lower, ``$-$,'' one to its negative branch.

A quantitative difference exists between Figure~\ref{Fig:DissipativeSituationSi}B and Figure~\ref{Fig:DissipativeSituationSi}D: The low-$\omega$ behavior of $\mathrm{Im}[k_{\mathrm{eff}}(\omega,L)]$ differs from one plot to the other. From Equations~\eqref{Eq:CharacteristicEquation-b} and \eqref{Eq:AdiabaticApproximation}, we easily check that
\begin{align}
\notag
\mathrm{Im}[k_{\mathrm{eff}}(\omega\to0,L)]&\left.\simeq\mathrm{Im}[k_{\mathrm{eff}}(\omega=0,L)]\right. \\
\label{Eq:ImaginaryPartsFromBelowAbove}
&\left.\hphantom{\simeq}\mp\mathrm{sign}(\beta_{2})\,\frac{|\beta_{2}|\,\gamma}{\alpha_{2}}\,\omega^{2}\right.
\end{align}
at the second order in $\omega\to0$, where ``$\mp$'' refers to the ``$\pm$'' branch of the Bogoliubov dispersion relation, and we recall that $\gamma$ and $\alpha_{2}$ are both positive (see Table~\ref{Tab:ExperimentalParameters}). In the TM-mode case of Figure~\ref{Fig:DissipativeSituationSi}B, $\mathrm{sign}(\beta_{2})=1$, as a result of which the imaginary part of the ``$+$'' (``$-$'') branch of the dispersion approaches quadratically its $\omega=0$ value from below (above). In the TE-mode case of Figure~\ref{Fig:DissipativeSituationSi}D, $\mathrm{sign}(\beta_{2})=-1$ and one then has the contrary behavior: The imaginary part of the ``$+$'' (``$-$'') branch of the dispersion tends quadratically to its $\omega=0$ value from above (below).

There is also a quantitative difference between Figure~\ref{Fig:DissipativeSituationSiNi}B and Figures~\ref{Fig:DissipativeSituationSi}B and \ref{Fig:DissipativeSituationSi}D: The $\omega=0$ value of $\mathrm{Im}[k_{\mathrm{eff}}(\omega,L)]$ equals its large-$\omega$ one in Figure~\ref{Fig:DissipativeSituationSiNi}B whereas it does not in Figures~\ref{Fig:DissipativeSituationSi}B and \ref{Fig:DissipativeSituationSi}D. This low-$\omega$ behavior distinctness originates from two-photon absorption that is negligible in silicon nitride while really present in silicon (see Table~\ref{Tab:ExperimentalParameters}). Indeed, from Equations~\eqref{Eq:CharacteristicEquation-b} and \eqref{Eq:AdiabaticApproximation}, it is easy to demonstrate that
\begin{align}
\notag
\mathrm{Im}[k_{\mathrm{eff}}(\omega=0,L)]&\left.=\mathrm{Im}[k_{\mathrm{eff}}(|\omega|\to\infty,L)]\right. \\
\label{Eq:GapedGaplessImaginaryParts}
&\left.\hphantom{=}\pm\frac{\alpha_{2}}{2}\,\langle\rho_{0}(z)\rangle_{L}.\right.
\end{align}
Thus, the $\omega=0$ and $|\omega|\to\infty$ imaginary parts of $k_{\mathrm{eff}}(\omega,L)$ are equal when $\alpha_{2}=0$, i.e., in the absence of two-photon absorption (silicon nitride), and different when $\alpha_{2}\neq0$, i.e., in the presence of two-photon absorption (silicon).

The Bogoliubov dispersion relation plotted in Figures~\ref{Fig:DissipativeSituationSiNi}C and \ref{Fig:DissipativeSituationSiNi}D is as for it different from the ones shown in Figures~\ref{Fig:DissipativeSituationSi}, \ref{Fig:DissipativeSituationSiNi}A, and \ref{Fig:DissipativeSituationSiNi}B, but its real part nevertheless looks like the lossless, stable result of Figure~\ref{Fig:DissipationlessSituation}A. This may be quantitatively investigated by making use of the $\alpha_{2}\to0$, i.e., the silicon-nitride, version of Equation~\eqref{Eq:CharacteristicEquation-b} and the adiabatic identification \eqref{Eq:AdiabaticApproximation}. In the present TE-mode case, $\beta_{2}$ and $\gamma$ are both positive (see Table~\ref{Tab:ExperimentalParameters}), as a result of which the square root in the $\alpha_{2}\to0$ local Bogoliubov law $k(\omega,z)$, Equation~\eqref{Eq:CharacteristicEquation-b}, is real for all $\omega$ and $\alpha_{0}/2$ is the $\omega$-independent imaginary part of $k(\omega,z)$. Thus, in virtue of \eqref{Eq:AdiabaticApproximation}, the space average over the segment $[0,L]$ of the first (second) row of $k(\omega,z)$ when $\alpha_{2}\to0$ corresponds to the imaginary (real) part of $k_{\mathrm{eff}}(\omega,L)$. Precisely,
\begin{equation}
\label{Eq:AdiabaticImaginaryPart}
\mathrm{Im}[k_{\mathrm{eff}}(\omega,L)]=\frac{\alpha_{0}}{2}\simeq2.30~\mathrm{m}^{-1}
\end{equation}
does not depend on $\omega$ and, defining the respective local, $z$-dependent, versions
\begin{align}
\label{Eq:LocalSpeedOfSound}
v^{-1}(z)&=\sqrt{\beta_{2}\,\gamma\,\rho_{0}(0)\,e^{-\alpha_{0}z}}, \\
\label{Eq:LocalHealingLength}
\tau(z)&=\sqrt{\frac{\beta_{2}}{\gamma\,\rho_{0}(0)\,e^{-\alpha_{0}z}}}=\frac{\beta_{2}}{v^{-1}(z)}
\end{align}
of Equations~\eqref{Eq:SpeedOfSound} and \eqref{Eq:HealingLength} for a $\rho_{0}(z)$ given by the $\alpha_{2}\to0$ limit of Equation~\eqref{Eq:MonochromaticDensity}, one finds the following low- and large-$\omega$ behaviors for $\mathrm{Re}[k_{\mathrm{eff}}(\omega,L)]$:
\begin{subequations}
\label{Eq:AdiabaticRealPart}
\begin{align}
\notag
&\left.\mathrm{Re}[k_{\mathrm{eff}}(\omega,L)]\right. \\
\label{Eq:AdiabaticRealPart-a}
&\left.\quad=\pm\left\langle\sqrt{\frac{\beta_{2}}{2}\,\omega^{2}\,\bigg[\frac{\beta_{2}}{2}\,\omega^{2}+2\,\gamma\,\rho_{0}(0)\,e^{-\alpha_{0}z}\bigg]}\right\rangle_{L}\right. \\
\label{Eq:AdiabaticRealPart-b}
&\left.\quad\simeq\pm
\begin{cases}
\langle v^{-1}(z)\rangle_{L}\,|\omega|, & |\omega|\ll1/\tau(L), \\
\displaystyle{\frac{\beta_{2}}{2}\,\omega^{2}+\gamma\,\langle\rho_{0}(0)\,e^{-\alpha_{0}z}\rangle_{L}}, & |\omega|\gg1/\tau(0),
\end{cases}
\right.
\end{align}
\end{subequations}
where $1/\tau(L)\simeq0.16~\mathrm{ps}^{-1}$, $1/\tau(0)\simeq0.26~\mathrm{ps}^{-1}$,
\begin{subequations}
\label{Eq:AdiabaticSpeedofSoundSiNi}
\begin{align}
\label{Eq:AdiabaticSpeedofSoundSiNi-a}
\langle v^{-1}(z)\rangle_{L}&=\frac{z_{\mathrm{eff}}(L/2)}{L/2}\,v^{-1}(0) \\
\label{Eq:AdiabaticSpeedofSoundSiNi-b}
&\simeq6.25\times10^{-2}~\mathrm{ps}\cdot\mathrm{m}^{-1},
\end{align}
\end{subequations}
and
\begin{subequations}
\label{Eq:AdiabaticNonlinearitySiNi}
\begin{align}
\label{Eq:AdiabaticNonlinearitySiNi-a}
\gamma\,\langle\rho_{0}(0)\,e^{-\alpha_{0}z}\rangle_{L}&=\frac{z_{\mathrm{eff}}(L)}{L}\,\gamma\,\rho_{0}(0) \\
\label{Eq:AdiabaticNonlinearitySiNi-b}
&\simeq1.32\times10^{-2}~\mathrm{m}^{-1},
\end{align}
\end{subequations}
$z_{\mathrm{eff}}(z)=(1-e^{-\alpha_{0}z})/\alpha_{0}\leqslant z$ denoting the effective length \cite{Agrawal1995} of a portion of waveguide of length $z\in[0,L]$. For lisibility's sake, we do not display the asymptotic approximations \eqref{Eq:AdiabaticRealPart-b} in Figure~\ref{Fig:DissipativeSituationSiNi}C.

\section{Proposed experiment}
\label{Sec:Experiment}

In this section, we propose an experiment \cite{Biasi2016} by means of which the Bogoliubov dispersion relation investigated in Section~\ref{Sec:BogoliubovDispersionRelation} can be measured. In Section~\ref{SubSec:ObservableToMeasure} first, we theoretically deal with the physical observable that has to be measured to get the Bogoliubov dispersion relation. In Section~\ref{SubSec:ExperimentalSetup} then, we present the basics of the experimental setup and how the theoretical ingredients of Section~\ref{SubSec:ObservableToMeasure} practically take part in the experiment.

\subsection{Observable to measure}
\label{SubSec:ObservableToMeasure}

As defined through Equation~\eqref{Eq:PlaneWaveSolution} (lossless configuration), Equation~\eqref{Eq:NearlyPlaneWaveSolution-b} (adiabatic lossy configuration), or Equation~\eqref{Eq:FormalMatrixExponentialForm} (generic, possibly nonadiabatic, lossy configuration), the Bogoliubov dispersion relation, denoted in each case as $k(\omega)$, $\langle k(\omega,z)\rangle_{z}$, and $k_{\mathrm{eff}}(\omega,z)$, is related to the phase of the $z$-dependent angular-frequency components $f_{\pm}(\omega,z)$ of the fluctuations $\varrho(t,z)$ and $\vartheta(t,z)$ of the power and the phase of the light beam in the waveguide. Then, it should be possible to extract it from a measurement of the phase $\phi_{L}(\omega)$ that a perturbation of the amplitude of the complex electric field accumulates during propagation along the waveguide. This accumulated phase is measured at $z=L$ as a function of the fluctuation's angular frequency $\omega$.

Considering that the amplitude $\bar{A}(t,z)$ of the in-air, $z\notin[0,L]$, complex optical field weakly deviates as
\begin{equation}
\label{Eq:InAirEnvelope}
\bar{A}(t,z)=\bar{A}_{0}(z)+e^{i\bar{\theta}_{0}(z)}\int_{-\infty}^{\infty}\frac{d\omega}{2\pi}\,\bar{a}(\omega,z)\,e^{-i\omega t}
\end{equation}
from the $t$-independent piecewise-constant mean field
\begin{equation}
\label{Eq:InAirMeanField}
\bar{A}_{0}(z)=\sqrt{\bar{\rho}_{0}(z)}\,e^{i\bar{\theta}_{0}(z)}=
\begin{cases}
\sqrt{\rho_{<}}\,e^{i\theta_{<}}, & z<0, \\
\sqrt{\rho_{>}}\,e^{i\theta_{>}}, & z>L,
\end{cases}
\end{equation}
the accumulated phase $\phi_{L}(\omega)$ introduced above should then read as
\begin{equation}
\label{Eq:AccumulatedPhase}
\phi_{L}(\omega)\equiv\mathrm{Arg}\bigg[\frac{\bar{a}(\omega,L^{+})}{\bar{a}(\omega,0^{-})}\bigg]+\mathrm{\Delta}\theta\pmod{2\pi},
\end{equation}
where $\mathrm{Arg}(X)\in\left]-\pi,\pi\right]$ denotes the principal argument of some complex number $X$ and $\mathrm{\Delta}\theta=\theta_{>}-\theta_{<}$. Here, the constants $\rho_{\lessgtr}$ and $\theta_{\lessgtr}$ correspond to the power and the phase of the in-air beam of light before (``$<$'') and after (``$>$'') propagation along the nonlinear waveguide. The Bogoliubov dispersion relation enters the formula \eqref{Eq:AccumulatedPhase} through the first term in the right-hand side, that is, through the relation linking the output Fourier component $\bar{a}(\omega,L^{+})$ of the in-air perturbation $\bar{A}(t,z)-\bar{A}_{0}(z)$ to its input, $\bar{a}(\omega,0^{-})$, one.

This relation may be deduced from the matching of the Poynting vector at the $z=0,L$ air-waveguide interfaces, that is, within the slowly-varying-envelope approximation used in this work, from the system \cite{Larre2016a}
\begin{subequations}
\label{Eq:MatchingPoyntingVector}
\begin{align}
\label{Eq:MatchingPoyntingVector-a}
\bar{A}(t,0^{-})&=\sqrt{n_{0}}\,A(t,0), \\
\label{Eq:MatchingPoyntingVector-b}
\sqrt{n_{0}}\,A(t,L)\,e^{i\beta_{0}L}&=\bar{A}(t,L^{+})\,e^{i\beta_{0}L/n_{0}}.
\end{align}
\end{subequations}
Linearizing the Madelung representation \eqref{Eq:MadelungRepresentation} of the amplitude $A(t,z)$ of the in-waveguide, $z\in[0,L]$, complex electric field according to Equations~\eqref{Eq:DensityExpansion} and \eqref{Eq:PhaseExpansion} yields, making use of Equations~\eqref{Eq:DensityFluctuation} and \eqref{Eq:PhaseFluctuation},
\begin{align}
\notag
A(t,z)&\left.=A_{0}(z)+e^{i\theta_{0}(z)}\int_{-\infty}^{\infty}\frac{d\omega}{2\pi}\,[a(\omega)\,u(\omega,z)\,e^{-i\omega t}\right. \\
\label{Eq:InWaveguideEnvelope}
&\left.\hphantom{=}+a^{\ast}(\omega)\,v^{\ast}(\omega,z)\,e^{i\omega t}],\right.
\end{align}
where $u(\omega,z)$ and $v(\omega,z)$, defined through \cite{Petrov2003, Petrov2004}
\begin{equation}
\label{Eq:BogoliubovAmplitudes}
u(\omega,z)\pm v(\omega,z)=f_{\pm}(\omega,z),
\end{equation}
are the Bogoliubov amplitudes \cite{Castin2001, Pethick2002, Pitaevskii2016}, as appearing in the context of dilute atomic Bose gases. In the absence of one- and two-photon losses, the non-Hermitianity of the Bogoliubov-de Gennes-type matrix \eqref{Eq:BogoliubovDeGennesEquations-b} makes the so-called Bogoliubov wavefunction $^{t}[u~v]$ obey the normalization condition $|u|^{2}-|v|^{2}=\pm\,1$ \cite{Castin2001}, where the ``$+$'' (``$-$'') sign refers to the ``positive'' (``negative'') branch of the Bogoliubov dispersion relation $k(\omega)$; since $|u|^{2}-|v|^{2}=\mathrm{Re}(f_{+}^{\ast}\,f_{-}^{\vphantom{\ast}})$ [from Equation~\eqref{Eq:BogoliubovAmplitudes}], this normalization constraint directly transfers to the $f_{\pm}$'s as $\mathrm{Re}(f_{+}^{\ast}\,f_{-}^{\vphantom{\ast}})=\pm\,1$. In the general case where one- and two-photon losses occur at the carrier angular frequency $\mathrm{\Omega}$, the Bogoliubov wavefunction obeys a related, yet formally more cumbersome [see Equation~\eqref{Eq:FormalMatrixExponentialForm}] normalization condition that one may generically write in the form
\begin{equation}
\label{Eq:NormalizationBogo}
|u(\omega,z)|^{2}-|v(\omega,z)|^{2}=N(\omega,z)\in\mathbb{R},
\end{equation}
where, according to the discussion above, $N(\omega,z)=\pm\,1$ for all $z$ as long as $\alpha_{0},\alpha_{2}=0$; from Equations~\eqref{Eq:BogoliubovAmplitudes} and \eqref{Eq:NormalizationBogo}, one has the normalization constraint $\mathrm{Re}[f_{+}^{\ast}(\omega,z)\,f_{-}^{\vphantom{\ast}}(\omega,z)]=N(\omega,z)$ for the $f_{\pm}(\omega,z)$'s in the presence of photonic losses. Combining Equations~\eqref{Eq:MonochromaticLongitudinalEnvelope}, \eqref{Eq:InAirEnvelope}, \eqref{Eq:InAirMeanField}, \eqref{Eq:MatchingPoyntingVector}, \eqref{Eq:InWaveguideEnvelope}, and \eqref{Eq:NormalizationBogo}, we eventually find
\begin{align}
\label{Eq:EntranceMeanField}
\rho_{<}&=n_{0}\,\rho_{0}(0),&\theta_{<}&=\theta_{0}(0), \\
\label{Eq:ExitMeanField}
\rho_{>}&=n_{0}\,\rho_{0}(L),&\theta_{>}&=\theta_{0}(L)+(1-1/n_{0})\,\beta_{0}\,L,
\end{align}
and, most importantly,
\begin{equation}
\label{Eq:InputOutputRelation-a}
\bar{a}(\omega,L^{+})=\frac{U(\omega)}{N(\omega,0)}\,\bar{a}(\omega,0^{-})+\frac{V^{\ast}(\omega)}{N(\omega,0)}\,\bar{a}^{\ast}(-\omega,0^{-}),
\end{equation}
where we have defined
\begin{align}
\label{Eq:InputOutputRelation-b}
U(\omega)&=u(\omega,L)\,u^{\ast}(\omega,0)-v^{\ast}(\omega,L)\,v(\omega,0), \\
\label{Eq:InputOutputRelation-c}
V(\omega)&=v(\omega,L)\,u^{\ast}(\omega,0)-u^{\ast}(\omega,L)\,v(\omega,0).
\end{align}

We now fix the input, $z=0^{-}$, condition for the perturbation $\bar{A}(t,z)-\bar{A}_{0}(z)$ on top of $\bar{A}_{0}(z)$ as
\begin{equation}
\label{Eq:InputConditionFluctuations}
\bar{a}(\omega,0^{-})\neq0\quad\text{while}\quad\bar{a}(-\omega,0^{-})=0,
\end{equation}
which physically amounts to consider that a single perturbation oscillating at $+\,\omega$ is injected into the waveguide, in accordance with the pump-and-probe experiment described in Section~\ref{SubSec:ExperimentalSetup}. As a result, according to Equation~\eqref{Eq:InputOutputRelation-a}, the formula \eqref{Eq:AccumulatedPhase} for the phase accumulated by the Bogoliubov fluctuations along the waveguide reduces to
\begin{equation}
\label{Eq:AccumulatedPhaseInputConditionFluctuations}
\phi_{L}(\omega)\equiv\mathrm{Arg}[U(\omega)]+\mathrm{\Delta}\theta\pmod{2\pi}.
\end{equation}
This congruence is strictly speaking valid in the case where $N(\omega,0)>0$; when $N(\omega,0)<0$, an extra $+\,\pi$ shift appears in the right-hand side but the latter may be absorbed in $\mathrm{\Delta}\theta$, as a result of which \eqref{Eq:AccumulatedPhaseInputConditionFluctuations} remains structurally valid also in the case where $N(\omega,0)<0$. From the generic diagonalization of Section~\ref{SubSubSec:ArbitraryEvolution} and making use of Equations~\eqref{Eq:BogoliubovAmplitudes}--\eqref{Eq:ExitMeanField} and \eqref{Eq:InputOutputRelation-b}, inverting Equation~\eqref{Eq:AccumulatedPhaseInputConditionFluctuations} should in principle yield the Bogoliubov dispersion relation. We illustrate this in Sections~\ref{SubSubSec:LosslessWaveguidePhase} and \ref{SubSubSec:LossyWaveguidePhase} in the previously-studied physically interesting cases where the (real part of the) Bogoliubov dispersion relation displays a linear, soundlike, behavior at low $\omega$, that is, in the lossless configuration of Figure~\ref{Fig:DissipationlessSituation}A and the lossy situation of Figures~\ref{Fig:DissipativeSituationSiNi}C and \ref{Fig:DissipativeSituationSiNi}D, respectively.

\subsubsection{Lossless waveguide}
\label{SubSubSec:LosslessWaveguidePhase}

In the lossless, $\alpha_{0},\alpha_{2}=0$, situation treated in Section~\ref{SubSec:LosslessWaveguide}, letting
\begin{align}
\label{Eq:DissipationlessBogoliubovAmplitudes-a}
u(\omega,z)&=\tilde{u}(\omega)\,e^{ik(\omega)z}, \\
\label{Eq:DissipationlessBogoliubovAmplitudes-b}
v(\omega,z)&=\tilde{v}(\omega)\,e^{ik(\omega)z},
\end{align}
Equation~\eqref{Eq:InputOutputRelation-b} transforms into
\begin{equation}
\label{Eq:DissipationlessInputOutputRelation-b}
U(\omega)=|\tilde{u}(\omega)|^{2}\,e^{ik(\omega)L}-|\tilde{v}(\omega)|^{2}\,e^{-ik^{\ast}(\omega)L}.
\end{equation}

\begin{figure*}[t!]
\includegraphics[width=\linewidth]{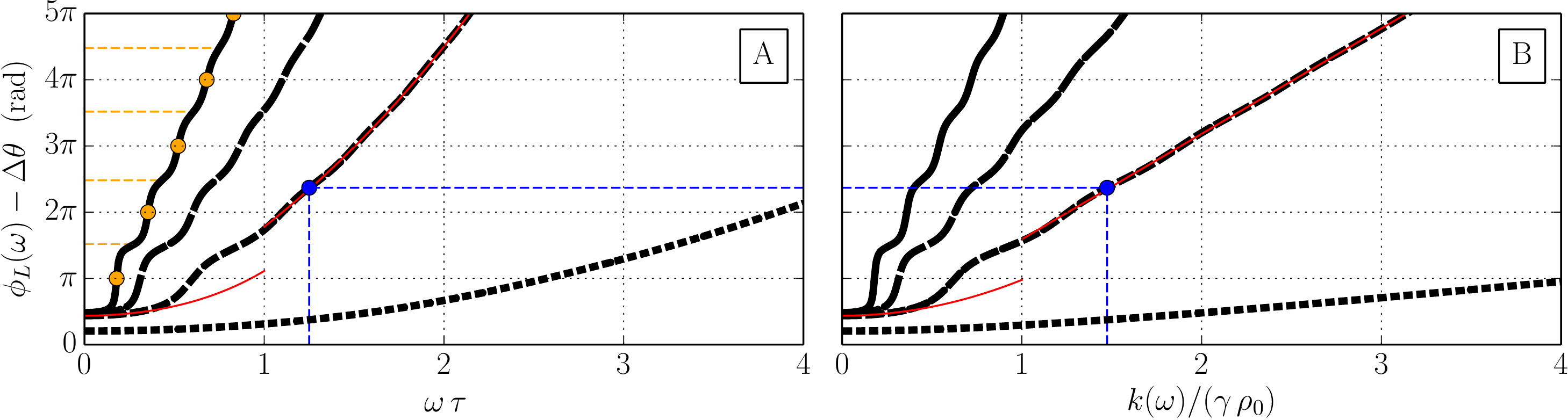}
\caption{Phase $\phi_{L}(\omega)$ accumulated by positive-branch [``$+$'' sign in Equation~\eqref{Eq:PlaneWaveWavenumber-b}; see the text] Bogoliubov fluctuations of the amplitude of the electric field in the course of its propagation along a lossless, $\alpha_{0},\alpha_{2}=0$, single-mode channel waveguide with $\beta_{2},\gamma>0$ and of normalized length $\ell=\gamma\,\rho_{0}\,L=0.75$ (black densely dashed curves), $5$ (dashed), $10$ (weakly dashed), and $17.5$ (plain). As pictorially sketched in blue, one first traces $\phi_{L}(\omega)$ as a function of the angular frequency $\omega$ of the Bogoliubov fluctuations [Panel A; from Equation~\eqref{Eq:BogoliubovWeights-a}], which then makes it possible to extract the wave number $k=k(\omega)$ of these Bogoliubov fluctuations, the so-called Bogoliubov dispersion relation of the fluid of light (here homogeneous of constant ``density'' $\rho_{0}$), making use of the plot of $\phi_{L}(\omega)$ against $k(\omega)$ [Panel B; from Equation~\eqref{Eq:BogoliubovWeights-b}]. On the $\ell=17.5$ curve of Panel A, the orange circular markers and the orange dashed lines indicate the inflection points of ordinates \eqref{Eq:InflectionOrdinates} and the plateaux \eqref{Eq:PlateauOrdinates-a}, respectively. The red curves correspond to the $\ell=5$ low- and large-$\omega$ behaviors \eqref{Eq:LowOmegaDissipationlessAccumulatedPhase} and \eqref{Eq:LargeOmegaDissipationlessAccumulatedPhase} where $k(\omega)/(\gamma\,\rho_{0})\simeq|\omega|\,\tau$ and $k(\omega)/(\gamma\,\rho_{0})\simeq\omega^{2}\,\tau^{2}/2+1$, respectively.}
\label{Fig:IdealAccumulatedPhase}
\end{figure*}

Considering for the sake of simplicity the positive, ``$+$,'' branch of $k(\omega)$ in Equation~\eqref{Eq:PlaneWaveWavenumber-b} and that the parameters $\beta_{2}$ and $\gamma$ entering it are both positive, $k(\omega)$ is positive for all $\omega$ and the Bogoliubov weights $\tilde{u}(\omega)$ and $\tilde{v}(\omega)$, such that $\tilde{u}(\omega)\pm\tilde{v}(\omega)=\tilde{f}_{\pm}(\omega)$ [see the definitions \eqref{Eq:PlaneWaveSolution} and \eqref{Eq:BogoliubovAmplitudes}], are real functions of $\omega$ satisfying
\begin{subequations}
\label{Eq:BogoliubovWeights}
\begin{align}
\label{Eq:BogoliubovWeights-a}
\tilde{u}(\omega)\pm\tilde{v}(\omega)&=\bigg(\frac{\omega^{2}\,\tau^{2}}{\omega^{2}\,\tau^{2}+4}\bigg)^{\pm\frac{1}{4}} \\
\label{Eq:BogoliubovWeights-b}
&=\bigg\{\frac{\sqrt{1+[k(\omega)/(\gamma\,\rho_{0})]^{2}}-1}{k(\omega)/(\gamma\,\rho_{0})}\bigg\}^{\pm\frac{1}{2}},
\end{align}
\end{subequations}
where $\tau$ is defined in Equation~\eqref{Eq:HealingLength}. From Equations~\eqref{Eq:PlaneWaveWavenumber-b}, \eqref{Eq:AccumulatedPhaseInputConditionFluctuations}, \eqref{Eq:DissipationlessInputOutputRelation-b}, and \eqref{Eq:BogoliubovWeights}, we plot in Figure~\ref{Fig:IdealAccumulatedPhase} the phase $\phi_{L}(\omega)-\mathrm{\Delta}\theta$ as a function of $\omega\,\tau\geqslant0$ [from Equation~\eqref{Eq:BogoliubovWeights-a}] and as a function of $k(\omega)/(\gamma\,\rho_{0})$ [from Equation~\eqref{Eq:BogoliubovWeights-b}].

In the low-$\omega$, $|\omega|\,\tau\ll1$, regime where $k(\omega)/(\gamma\,\rho_{0})\simeq|\omega|\,\tau$ [first row of Equation~\eqref{Eq:PlaneWaveWavenumberLimits}], a lengthy Taylor expansion of Equation~\eqref{Eq:AccumulatedPhaseInputConditionFluctuations} yields
\begin{align}
\notag
\phi_{L}(\omega)-\mathrm{\Delta}\theta&\left.=2\arctan\!\bigg(\frac{\ell}{1+\sqrt{1+\ell^{2}}}\bigg)\right. \\
\label{Eq:LowOmegaDissipationlessAccumulatedPhase}
&\left.\hphantom{\simeq}+\frac{\ell\,(3+2\,\ell^{2})}{6\,(1+\ell^{2})}\,\bigg[\frac{k(\omega)}{\gamma\,\rho_{0}}\bigg]^{2}+{\cdots},\right.
\end{align}
where $\ell=\gamma\,\rho_{0}\,L$ is the waveguide's length in units of the ``nonlinear length'' $1/(\gamma\,\rho_{0})$. The approximation \eqref{Eq:LowOmegaDissipationlessAccumulatedPhase} straightforwardly reduces to
\begin{equation}
\label{Eq:LowOmegaDissipationlessAccumulatedPhaseBis}
\phi_{L}(\omega)-\mathrm{\Delta}\theta\simeq\ell+\frac{\ell}{2}\,\bigg[\frac{k(\omega)}{\gamma\,\rho_{0}}\bigg]^{2}+\cdots
\end{equation}
in the particular limit $\ell\ll1$, and to
\begin{equation}
\label{Eq:LowOmegaDissipationlessAccumulatedPhaseBisBis}
\phi_{L}(\omega)-\mathrm{\Delta}\theta\simeq\frac{\pi}{2}+\frac{\ell}{3}\,\bigg[\frac{k(\omega)}{\gamma\,\rho_{0}}\bigg]^{2}+\cdots
\end{equation}
when $\ell\gg1$. The latter approximations are all the more satisfied as the second term is much smaller than the first one in each right-hand side, i.e., as $|\omega|\,\tau\ll1/\sqrt{\ell}$; consequently, \eqref{Eq:LowOmegaDissipationlessAccumulatedPhaseBis} and \eqref{Eq:LowOmegaDissipationlessAccumulatedPhaseBisBis} are valid when $|\omega|\,\tau\ll1\ll1/\sqrt{\ell}$ and when $|\omega|\,\tau\ll1/\sqrt{\ell}\ll1$, respectively. Importantly, as one sees in Equations~\eqref{Eq:LowOmegaDissipationlessAccumulatedPhase}--\eqref{Eq:LowOmegaDissipationlessAccumulatedPhaseBisBis}, the low-$\omega$ Bogoliubov dispersion relation $k(\omega)\simeq v^{-1}\,|\omega|$ may be extracted from the phase $\phi_{L}(\omega)-\mathrm{\Delta}\theta$ by Taylor expanding the latter at the second order---at least---in $|\omega|\,\tau\ll1$.

At the angular frequencies $\omega\geqslant0$ such that
\begin{align}
\label{Eq:InflectionAbscissas-a}
\mathrm{Arg}[U(\omega^{\mp})]&=\pm\,\pi^{\mp}\quad\text{or} \\
\label{Eq:InflectionAbscissas-b}
\mathrm{Arg}[U(\omega)]&=0,
\end{align}
the graph of $\phi_{L}(\omega)-\mathrm{\Delta}\theta$ presents inflection points of ordinates
\begin{equation}
\label{Eq:InflectionOrdinates}
[\phi_{L}(\omega)-\mathrm{\Delta}\theta]_{n}=n\,\pi,
\end{equation}
where $n\in\mathbb{N}^{\ast}$. On the other hand, in between two successive inflection points, $\phi_{L}(\omega)-\mathrm{\Delta}\theta$ smoothly varies around the discrete plateaux
\begin{subequations}
\label{Eq:PlateauOrdinates}
\begin{align}
\label{Eq:PlateauOrdinates-a}
[\phi_{L}(\omega)-\mathrm{\Delta}\theta]_{n'}&=2\,n'\,\pi\mp2\arctan\!\bigg(\frac{\ell}{1+\sqrt{1+\ell^{2}}}\bigg) \\
\label{Eq:PlateauOrdinates-b}
&\simeq
\begin{cases}
2\,n'\,\pi\mp\ell, & \ell\ll1, \\
2\,n'\,\pi\mp\pi/2, & \ell\gg1,
\end{cases}
\end{align}
\end{subequations}
where $n'\in\mathbb{N}^{\ast}$. This explains the smooth staircase structures observed in Figure~\ref{Fig:IdealAccumulatedPhase}. When $\ell\ll1$, one shows that the points of ordinates \eqref{Eq:PlateauOrdinates-a} almost coincide with the inflection points of ordinates \eqref{Eq:InflectionOrdinates} with $n=2\,n'$, as one notes (for the ordinates) in the first row of Equation~\eqref{Eq:PlateauOrdinates-b}; in this case, the staircase features disappear, as examplified by the $\ell=0.75$ curves of Figure~\ref{Fig:IdealAccumulatedPhase}. When $\ell\gg1$, the plateaux \eqref{Eq:PlateauOrdinates-a} are on the contrary very distinct from \eqref{Eq:InflectionOrdinates}, as shown in the second row of Equation~\eqref{Eq:PlateauOrdinates-b} and illustrated by, e.g., the $\ell=17.5$ curves of Figure~\ref{Fig:IdealAccumulatedPhase}.

In the large-$\omega$, $|\omega|\,\tau\gg1$, regime where $k(\omega)/(\gamma\,\rho_{0})\simeq\omega^{2}\,\tau^{2}/2+1$ [second row of Equation~\eqref{Eq:PlaneWaveWavenumberLimits}], one has from Equations~\eqref{Eq:BogoliubovWeights} the zeroth-order approximations $\tilde{u}(\omega)\simeq1$ and $\tilde{v}(\omega)\simeq0$, which leads to the very simple approximation
\begin{equation}
\label{Eq:LargeOmegaDissipationlessAccumulatedPhase}
\phi_{L}(\omega)-\mathrm{\Delta}\theta\simeq k(\omega)\,L=\frac{k(\omega)}{\gamma\,\rho_{0}}\,\ell,
\end{equation}
all the more satisfied as the right-hand side is large, i.e., as $|\omega|\,\tau\gg1/\sqrt{\ell}$. From Equation~\eqref{Eq:LargeOmegaDissipationlessAccumulatedPhase}, it is very easy to extract the large-$\omega$ Bogoliubov dispersion relation $k(\omega)$ of the uniform fluid of light.

This discussion shines interesting new light on the theoretical interpretation of the experiment of Reference~\cite{Vocke2015}, where the Bogoliubov dispersion relation in a waveguide-free paraxial-propagation geometry was directly extracted from the transmission phase. For the sake of uniformity with the rest of the paper, we carry out this discussion in terms of $k(\omega)$, but a translation to the situation of Reference~\cite{Vocke2015} is straightforward (see the second paragraph of Section~\ref{Sec:Conclusion}). In the large-$\omega$ limit where $\tilde{u}(\omega)\simeq1$ and $\tilde{v}(\omega)\simeq0$, the Bogoliubov dispersion relation is mostly particlelike and the plateau structure gives a negligible correction to $\phi_{L}(\omega)$. As one can see in Figure~\ref{Fig:IdealAccumulatedPhase}, the situation is different at lower $\omega$'s where the plateaux are very pronounced and may introduce dramatic deviations from the simple approximation \eqref{Eq:LargeOmegaDissipationlessAccumulatedPhase}.

While the experiment \cite{Vocke2015} could not access the deep sonic regime where the correction is the most important, still the presence of plateaux may explain the slight deviation between experiments and theoretical expectations. In any case, it is immediate to see from the analytical expression \eqref{Eq:DissipationlessInputOutputRelation-b} and from Figure~\ref{Fig:IdealAccumulatedPhase} that the coarse-grained shape of $\phi_{L}(\omega)$ when the plateau structure is smoothened out recovers the Bogoliubov dispersion $k(\omega)$ for almost all the $\omega$'s, except in the vicinity of $\omega=0$ where the first plateau remains.

\subsubsection{Lossy waveguide}
\label{SubSubSec:LossyWaveguidePhase}

In the lossy, $\alpha_{0},\alpha_{2}\neq0$, situation treated in Section~\ref{SubSec:LossyWaveguide}, letting
\begin{align}
\label{Eq:DissipativeBogoliubovAmplitudes-a}
u(\omega,z)&=\tilde{u}(\omega,z)\,e^{ik_{\mathrm{eff}}(\omega,z)z}, \\
\label{Eq:DissipativeBogoliubovAmplitudes-b}
v(\omega,z)&=\tilde{v}(\omega,z)\,e^{ik_{\mathrm{eff}}(\omega,z)z},
\end{align}
Equation~\eqref{Eq:InputOutputRelation-b} transforms into
\begin{align}
\notag
U(\omega)&\left.=\tilde{u}(\omega,L)\,\tilde{u}^{\ast}(\omega,0)\,e^{ik_{\mathrm{eff}}^{\vphantom{\ast}}(\omega,L)L}\right. \\
\label{Eq:DissipativeInputOutputRelation-b}
&\left.\hphantom{=}-\tilde{v}^{\ast}(\omega,L)\,\tilde{v}(\omega,0)\,e^{-ik_{\mathrm{eff}}^{\ast}(\omega,L)L}.\right.
\end{align}

\begin{figure*}[t!]
\includegraphics[width=\linewidth]{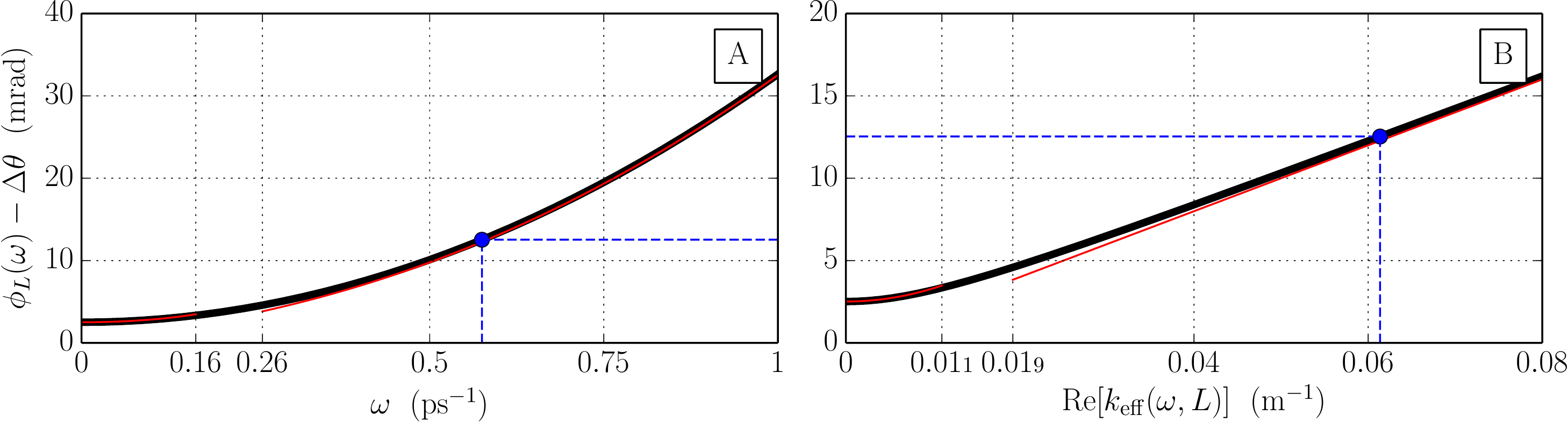}
\caption{Same as Figure~\ref{Fig:IdealAccumulatedPhase} for a TE mode propagating along a $L=20~\mathrm{cm}$-long single-mode channel waveguide with a silicon-nitride core. The operating wavelength equals $1.55~\mu\mathrm{m}$, the incident power is of $100~\mathrm{mW}$, and the corresponding waveguide's parameters are given in the right column of Table~\ref{Tab:ExperimentalParameters}. Accordingly, the normalized length \eqref{Eq:RevolutionRevolution} is approximately equal to $2.51\times10^{-3}$ and is then too small to observe the plateau structures of Figure~\ref{Fig:IdealAccumulatedPhase}. The red curves represent the $|\omega|\ll1/\tau(L)\simeq0.16~\mathrm{ps}^{-1}$ and $|\omega|\gg1/\tau(0)\simeq0.26~\mathrm{ps}^{-1}$ approximations \eqref{Eq:LowOmegaDissipativeAccumulatedPhase} and \eqref{Eq:LargeOmegaDissipativeAccumulatedPhase} where, respectively, $\mathrm{Re}[k_{\mathrm{eff}}(\omega,L)]\ll1.10\times10^{-2}~\mathrm{m}^{-1}$ is approximately phononlike, given by the first row of Equation~\eqref{Eq:AdiabaticRealPart-b}, and $\mathrm{Re}[k_{\mathrm{eff}}(\omega,L)]\gg1.92\times10^{-2}~\mathrm{m}^{-1}$ is asymptotically particlelike, given by the second row of Equation~\eqref{Eq:AdiabaticRealPart-b}.}
\label{Fig:RealisticAccumulatedPhase}
\end{figure*}

Now, we specifically focus on the TE-mode silicon-nitride case of Figures~\ref{Fig:DissipativeSituationSiNi}C and \ref{Fig:DissipativeSituationSiNi}D, for which the real part of $k_{\mathrm{eff}}(\omega,z=L)$ displays an interesting phononlike behavior at low angular frequency $\omega$ [cf.~upper row of Equation~\eqref{Eq:AdiabaticRealPart-b}]. Considering for simplicity's sake the positive, ``$+$,'' branch of $k_{\mathrm{eff}}(\omega,z)$ and as $\beta_{2}>0$ and $\gamma>0$ (cf.~right column of Table~\ref{Tab:ExperimentalParameters}), $\mathrm{Re}[k_{\mathrm{eff}}(\omega,z)]$ is positive and the $\tilde{u}(\omega,z)$'s and the $\tilde{v}(\omega,z)$'s, such that $\tilde{u}(\omega,z)\pm\tilde{v}(\omega,z)=\tilde{f}_{\pm}(\omega,z)$ [we make use of the adiabatic result \eqref{Eq:NearlyPlaneWaveSolution-b}, identifying $k_{\mathrm{eff}}(\omega,z)$ to $\langle k(\omega,z)\rangle_{z}$ as in Section~\ref{SubSubSec:ArbitraryEvolution}] are real quantities verifying
\begin{equation}
\label{Eq:LossyBogoliubovWeights}
\tilde{u}(\omega,z)\pm\tilde{v}(\omega,z)=\sqrt{N(\omega,z)}\,\bigg[\frac{\omega^{2}\,\tau^{2}(z)}{\omega^{2}\,\tau^{2}(z)+4}\bigg]^{\pm\frac{1}{4}},
\end{equation}
where the local time parameter $\tau(z)$ is defined in Equation~\eqref{Eq:LocalHealingLength} and $N(\omega,z)$ is here positive. Using Equations~\eqref{Eq:AdiabaticImaginaryPart}, \eqref{Eq:AdiabaticRealPart-a}, \eqref{Eq:AccumulatedPhaseInputConditionFluctuations}, \eqref{Eq:DissipativeInputOutputRelation-b}, and \eqref{Eq:LossyBogoliubovWeights}, we plot in Figure~\ref{Fig:RealisticAccumulatedPhase} the phase $\phi_{L}(\omega)-\mathrm{\Delta}\theta$ as a function of $\omega\geqslant0$ and as a function of $\mathrm{Re}[k_{\mathrm{eff}}(\omega,L)]$.

In the low-$\omega$, i.e., $|\omega|\ll1/\tau(L)\simeq0.16~\mathrm{ps}^{-1}$, regime where $\mathrm{Re}[k_{\mathrm{eff}}(\omega,L)]\ll1.10\times10^{-2}~\mathrm{m}^{-1}$ is soundlike, approximately given by the first row of Equation~\eqref{Eq:AdiabaticRealPart-b}, $\phi_{L}(\omega)-\mathrm{\Delta}\theta$ obeys a Taylor expansion similar to Equation~\eqref{Eq:LowOmegaDissipationlessAccumulatedPhase}:
\begin{subequations}
\label{Eq:LowOmegaDissipativeAccumulatedPhase}
\begin{align}
\notag
&\left.\phi_{L}(\omega)-\mathrm{\Delta}\theta\right. \\
\notag
&\left.\quad=2\arctan\!\bigg(\frac{\ell}{1+\sqrt{1+\ell^{2}}}\bigg)\right. \\
\label{Eq:LowOmegaDissipativeAccumulatedPhase-a}
&\left.\quad\hphantom{=}+\frac{\ell\,(3+2\,\ell^{2})}{6\,(1+\ell^{2})}\,\bigg\{\frac{\mathrm{Re}[k_{\mathrm{eff}}(\omega,L)]}{\mathcal{C}(\alpha_{0})\,\gamma\,\langle\rho_{0}(0)\,e^{-\alpha_{0}z}\rangle_{L}}\bigg\}^{2}+\cdots\right. \\
\label{Eq:LowOmegaDissipativeAccumulatedPhase-b}
&\left.\quad\simeq2.51\times10^{-3}+7.95\;\mathrm{Re}^{2}[k_{\mathrm{eff}}(\omega,L)]+{\cdots},\right.
\end{align}
\end{subequations}
where the parameters
\begin{align}
\label{Eq:Revolution}
\mathcal{C}(\alpha_{0})&=\frac{1}{\cosh^{2}(\alpha_{0}\,L/4)}\simeq0.95, \\
\label{Eq:RevolutionRevolution}
\ell&=\mathcal{C}(\alpha_{0})\,\gamma\,\langle\rho_{0}(0)\,e^{-\alpha_{0}z}\rangle_{L}\,L\simeq2.51\times10^{-3},
\end{align}
and $\mathrm{Re}[k_{\mathrm{eff}}(\omega,L)]$ is expressed in $\mathrm{m}^{-1}$. Note that as $\mathcal{C}(0)=1$, the expansion \eqref{Eq:LowOmegaDissipativeAccumulatedPhase-a} reduces to the lossless result \eqref{Eq:LowOmegaDissipationlessAccumulatedPhase} when $\alpha_{0}\to0$.

Contrary to the graphs plotted in Figure~\ref{Fig:IdealAccumulatedPhase}, the phase $\phi_{L}(\omega)-\mathrm{\Delta}\theta$ in Figure~\ref{Fig:RealisticAccumulatedPhase} displays no staircase feature. Following the fourth paragraph of Section~\ref{SubSubSec:LosslessWaveguidePhase}, this may be explained by the fact that the normalized length $\ell$ is in the present case very small, of the order of $10^{-3}$.

In the large-$\omega$, i.e., $|\omega|\gg1/\tau(0)\simeq0.26~\mathrm{ps}^{-1}$, regime where $\mathrm{Re}[k_{\mathrm{eff}}(\omega,L)]\gg1.92\times10^{-2}~\mathrm{m}^{-1}$ is particlelike, approximately given by the second row of Equation~\eqref{Eq:AdiabaticRealPart-b}, $\tilde{u}(\omega,z)\simeq\sqrt{N(\omega,z)}$ and $\tilde{v}(\omega,z)\simeq0$, as a result of which $\phi_{L}(\omega)-\mathrm{\Delta}\theta$ reduces to
\begin{equation}
\label{Eq:LargeOmegaDissipativeAccumulatedPhase}
\phi_{L}(\omega)-\mathrm{\Delta}\theta\simeq\mathrm{Re}[k_{\mathrm{eff}}(\omega,L)]\,L=0.20\;\mathrm{Re}[k_{\mathrm{eff}}(\omega,L)],
\end{equation}
where $\mathrm{Re}[k_{\mathrm{eff}}(\omega,L)]$ is once more expressed in $\mathrm{m}^{-1}$.

\subsection{Experimental setup}
\label{SubSec:ExperimentalSetup}

\begin{figure*}[t!]
\centering
\includegraphics[scale=0.9]{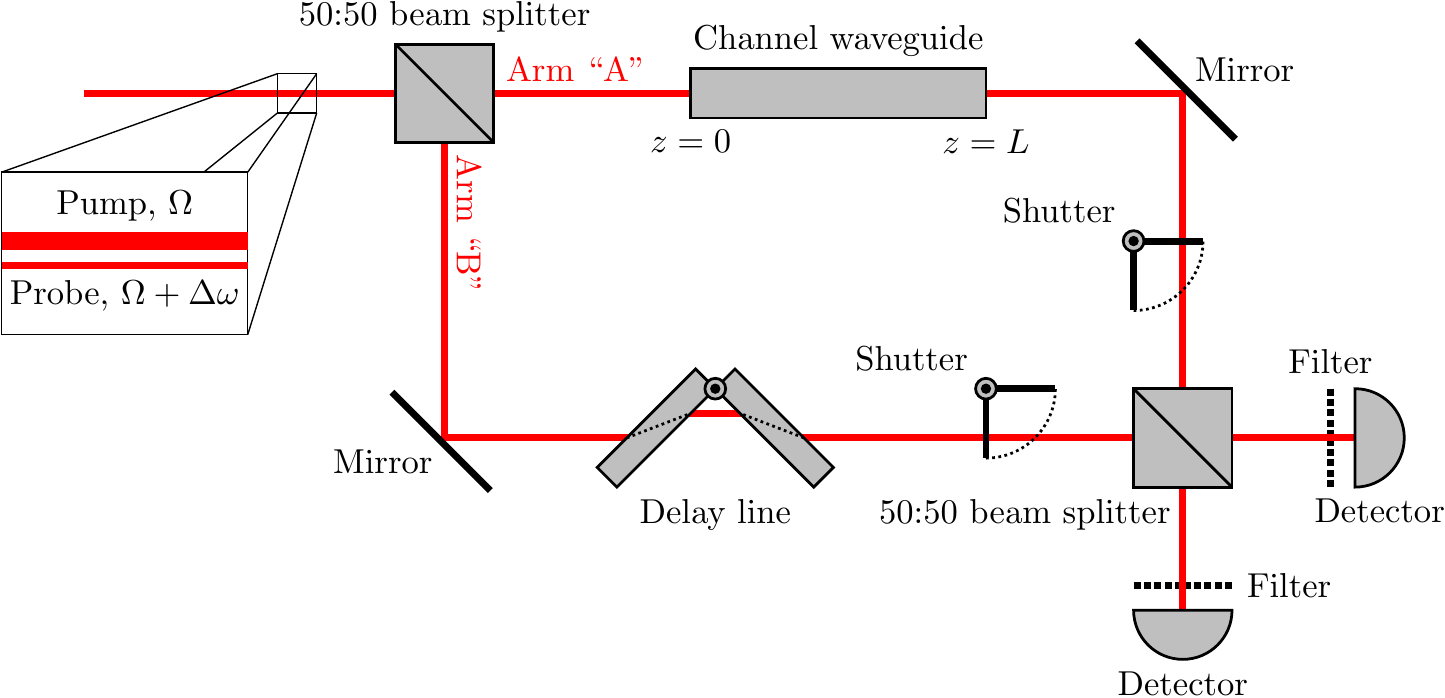}
\caption{Schematic representation of the Mach-Zehnder-interferometry pump-and-probe experiment \cite{Biasi2016} making it possible to measure \eqref{Eq:AccumulatedPhase}, \eqref{Eq:AccumulatedPhaseProbe} and then, as explained in Section~\ref{SubSec:ObservableToMeasure}, the Bogoliubov dispersion relation of the fluid of light propagating along the channel waveguide encompassed in between $z=0$ and $z=L$.}
\label{Fig:ExperimentalSetup}
\end{figure*}

The experimental setup \cite{Biasi2016} that we propose to measure the accumulated phase \eqref{Eq:AccumulatedPhase} and in turn, as detailed in Section~\ref{SubSec:ObservableToMeasure}, the Bogoliubov dispersion relation of the beam of light propagating along the channel waveguide is sketched in Figure~\ref{Fig:ExperimentalSetup}. It basically consists in a free-space Mach-Zehnder interferometer \cite{Hariharan2007} illuminated by a large-power pump beam of angular frequency $\mathrm{\Omega}$ and a collinear low-power probe beam of angular frequency $\mathrm{\Omega}+\mathrm{\Delta}\omega$ (with $\mathrm{\Delta}\omega\gtrless0$). One of the two arms of the Mach-Zehnder interferometer, denoted ``A,'' is focused on the channel waveguide encompassed in between $z=0$ and $z=L$ while the other one, denoted ``B,'' is free. The high- and low-power beams nonlinearly interact in the waveguide through a four-wave mixing. The total intensity $I_{\mathrm{p}}$ of the probe ``p'' measured at one of the light detectors of the Mach-Zehnder interferometer after filtering out all other frequency components (namely, the pump at $\mathrm{\Omega}$ and the Kerr-induced idler at $\mathrm{\Omega}-\mathrm{\Delta}\omega$) reads
\begin{equation}
\label{Eq:TotalIntensity}
I_{\mathrm{p}}=I_{\mathrm{A}}+I_{\mathrm{B}}+2\,\sqrt{I_{\mathrm{A}}\,I_{\mathrm{B}}}\cos[\phi_{L}(\mathrm{\Delta}\omega)+\cdots],
\end{equation}
where $I_{\mathrm{A}}$ ($I_{\mathrm{B}}$) is the intensity measured in the arm ``A'' (``B'') by switching off the arm ``B'' (``A'') by means of an optical shutter and $\phi_{L}(\mathrm{\Delta}\omega)+\cdots$ denotes the dephasing between the arm ``A'' and the arm ``B,'' induced most particularly by the presence of the waveguide along the arm ``A.'' Making use of a well-adjusted delay line for making the interferometer perfectly balanced \cite{Biasi2016}, the latter dephasing reduces to the phase
\begin{equation}
\label{Eq:AccumulatedPhaseProbe}
\phi_{L}(\mathrm{\Delta}\omega)\equiv\mathrm{Arg}\bigg[\frac{A_{\mathrm{p}}(\mathrm{\Delta}\omega,L^{+})}{A_{\mathrm{p}}(\mathrm{\Delta}\omega,0^{-})}\bigg]\pmod{2\pi}
\end{equation}
accumulated by the probe in the course of its propagation along the waveguide, hence the use of the suspension dots in Equation~\eqref{Eq:TotalIntensity}.

The notations used in Equation~\eqref{Eq:AccumulatedPhaseProbe} are identical to the ones used in Equation~\eqref{Eq:AccumulatedPhase} for the simple and good reason that the quantities \eqref{Eq:AccumulatedPhaseProbe} and \eqref{Eq:AccumulatedPhase} are strictly equal: The weak-power probe on top of the strong-power pump in the zoomed window of Figure~\ref{Fig:ExperimentalSetup} corresponds to the weak-amplitude fluctuation superimposing upon the steady profile $\bar{A}_{0}(z)$ in Equation~\eqref{Eq:InAirEnvelope}. This can be easily seen by defining
\begin{subequations}
\label{Eq:ProbeWave}
\begin{align}
\label{Eq:ProbeWave-a}
\bar{a}(\omega,z)&=2\pi\,\delta(\omega-\mathrm{\Delta}\omega)\,A_{\mathrm{p}}(\omega,z)\,e^{-i\bar{\theta}_{0}(z)}, \\
\label{Eq:ProbeWave-b}
A_{\mathrm{p}}(\omega,z)&=\tilde{A}_{\mathrm{p}}(\omega,z)\,e^{i\mathrm{\Delta}\beta_{\mathrm{p}}(\omega)z}
\end{align}
\end{subequations}
in Equation~\eqref{Eq:InAirEnvelope}, that indeed yields the usual decomposition
\begin{equation}
\label{Eq:PumpAndProbe}
\bar{A}(t,z)=\bar{A}_{0}(z)+\tilde{A}_{\mathrm{p}}(\mathrm{\Delta}\omega,z)\,e^{-i\mathrm{\Delta}\omega t}\,e^{i\mathrm{\Delta}\beta_{\mathrm{p}}(\mathrm{\Delta}\omega)z}
\end{equation}
for the total complex optical field's amplitude in terms of the pump, oscillating at $(\mathrm{\Omega},\beta_{0}/n_{0})$, and the probe, detuned from the former at $[\mathrm{\Omega}+\mathrm{\Delta}\omega,\beta_{0}/n_{0}+\mathrm{\Delta}\beta_{\mathrm{p}}(\mathrm{\Delta}\omega)]$. Similarly, the Bogoliubov wave on top of the stationary mean field $A_{0}(z)$ in Equation~\eqref{Eq:InWaveguideEnvelope} corresponds to the linear superposition of the signal ``s'' at $[\mathrm{\Omega}+\mathrm{\Delta}\omega,\beta_{0}+\mathrm{\Delta}\beta_{\mathrm{s}}(\mathrm{\Delta}\omega)]$ and the idler ``i'' at $[\mathrm{\Omega}-\mathrm{\Delta}\omega,\beta_{0}+\mathrm{\Delta}\beta_{\mathrm{i}}(\mathrm{\Delta}\omega)]$:
\begin{align}
\notag
A(t,z)&\left.=A_{0}(z)+\tilde{A}_{\mathrm{s}}(\mathrm{\Delta}\omega,z)\,e^{-i\mathrm{\Delta}\omega t}\,e^{i\mathrm{\Delta}\beta_{\mathrm{s}}(\mathrm{\Delta}\omega)z}\right. \\
\label{Eq:PumpSignalAndIdler}
&\left.\hphantom{=}+\tilde{A}_{\mathrm{i}}(\mathrm{\Delta}\omega,z)\,e^{i\mathrm{\Delta}\omega t}\,e^{i\mathrm{\Delta}\beta_{\mathrm{i}}(\mathrm{\Delta}\omega)z},\right.
\end{align}
where the signal's and idler's amplitudes $\tilde{A}_{\mathrm{s}}(\mathrm{\Delta}\omega,z)$ and $\tilde{A}_{\mathrm{i}}(\mathrm{\Delta}\omega,z)$ are defined through
\begin{subequations}
\label{Eq:SignalIdler}
\begin{align}
\label{Eq:SignalIdler-a}
\begin{bmatrix}
a(\omega)\,u(\omega,z) \\ a^{\ast}(\omega)\,v^{\ast}(\omega,z)
\end{bmatrix}
&=2\pi\,\delta(\omega-\mathrm{\Delta}\omega)
\begin{bmatrix}
A_{\mathrm{s}}(\omega,z) \\ A_{\mathrm{i}}(\omega,z)
\end{bmatrix}
e^{-i\theta_{0}(z)}, \\
\label{Eq:SignalIdler-b}
A_{\mathrm{s},\mathrm{i}}(\omega,z)&=\tilde{A}_{\mathrm{s},\mathrm{i}}(\omega,z)\,e^{i\mathrm{\Delta}\beta_{\mathrm{s},\mathrm{i}}(\omega)z}.
\end{align}
\end{subequations}

In this Mach-Zehnder-interferometry pump-and-probe experiment, one measures the probe dephasing $\phi_{L}(\mathrm{\Delta}\omega)$ as a function of the angular-frequency detuning $\mathrm{\Delta}\omega$---i.e., the angular frequency $\omega$ of the Bogoliubov fluctuation, from the $\delta$ peak in the definition \eqref{Eq:SignalIdler-a}---, from which one deduces the Bogoliubov dispersion relation of the fluid of light as a function of $\mathrm{\Delta}\omega$, making use of the recipe detailed in Section~\ref{SubSec:ObservableToMeasure}.

\section{Conclusion}
\label{Sec:Conclusion}

Making use of Bogoliubov's theory of dilute Bose quantum fluids, we have investigated the dispersion relation of small luminous fluctuations on top of a beam of polarized monochromatic light propagating along a single-mode channel waveguide displaying an instantaneous and spatially local Kerr nonlinearity as well as one- and two-photon losses. Analytical and numerical results have been derived in both the ideal situation where the photonic losses are absent and the realistic case where they are present. Two types of nonlinear-silicon-photonics waveguides with silicon and silicon-nitride cores have been used to illustrate our theoretical predictions. Additionally, we have proposed a Mach-Zehnder-interferometry pump-and-probe experiment \cite{Biasi2016} to measure the dispersion law of the Bogoliubov excitations of the beam of light: A weak-power probe beam (the analogous Bogoliubov wave) copropagates along the waveguide with a strong-power pump beam (the analogous background Bose quantum fluid) and accumulates a phase delay in the course of its propagation, from which the Bogoliubov dispersion relation is derived.

Importantly, note that our one-dimensional results in the time domain are conceptually very general and may be transfered (modulo \textit{ad-hoc} changes in the notations) to the full three-dimensional generalized nonlinear Schr\"odinger problem (see, e.g., References~\cite{Rosanov2002, Larre2015b} for the $a_{0},a_{2}=0$ situation)
\begin{align}
\notag
i\,\frac{\partial E}{\partial z}&\left.=-\frac{1}{2\,\beta_{0}}\,\bigg(\frac{\partial^{2}E}{\partial x^{2}}+\frac{\partial^{2}E}{\partial y^{2}}\bigg)+\frac{\beta_{2}}{2}\,\frac{\partial^{2}E}{\partial t^{2}}\right. \\
\label{Eq:FullNLSE}
&\left.\hphantom{=}-g\,|E|^{2}\,E-\frac{i}{2}\,(a_{0}+a_{2}\,|E|^{2})\,E,\right.
\end{align}
that describes the propagation of the slowly varying envelope of the total complex electric field of a paraxial beam of quasimonochromatic light in a waveguide-free, local, and lossy Kerr medium. In addition to its direct interest for nonlinear optics as a tool to probe novel effects in the optical phase, the experiment we propose holds the promise to highlight a very general feature of the generalized nonlinear Schr\"odinger equation.

\begin{acknowledgement}
We gratefully thank Santanu Manna and Fabio Turri for their collaboration in the early stages of this theoretical study as well as Quentin Glorieux and Arnaud Mussot for useful inputs, valuable comments, and interesting discussions. This work was financially supported by the Centre National de la Recherche Scientifique (CNRS), by the Provincia Autonoma di Trento through the Call ``Grandi Progetti 2012,'' Project ``On Silicon Chip Quantum Optics for Quantum Computing and Secure Communications---SiQuro,'' and by the European ``Future and Emerging Technologies'' Proactive Grant ``Analog Quantum Simulators for Many-Body Dynamics---AQuS,'' Project No.~640800.
\end{acknowledgement}

\section*{Author contribution statement}

All the authors contributed to this work. P.-\'E.~L.~did the calculations, the figures, and wrote the text. S.~B.~and F.~R.-M.~ran a numerical code providing the realistic figures listed in Tab.~\ref{Tab:ExperimentalParameters}, contributed to the text, and are presently working on the proposed experiment. L.~P.~contributed to the text. I.~C.~was in charge of the general supervision of the work and contributed to the text.

\appendix

\numberwithin{equation}{section}

\section{Adiabatic theorem for \texorpdfstring{$\boldsymbol{z}$}{Lg}-dependent propagating optical systems}
\label{App:AdiabaticTheoremForZDependentPropagatingOpticalSystems}

In this appendix, we reformulate the adiabatic theorem of quantum mechanics \cite{Born1928, Messiah1999} within the optical language. The derivation of the final result \eqref{Eq:AdiabaticEigensolution} is illustratively made in the standard case of a Hermitian effective evolution along the propagation, $z$, axis but there exists a similar identity in the non-Hermitian case (see, e.g., Reference~\cite{Sun1993}).

Let us consider that the propagation in the positive-$z$ direction of some (scalar or vector) complex optical field $\mathrm{\Psi}(z\in[0,L])$ is ruled by the generic Schr\"odinger-type equation
\begin{equation}
\label{Eq:TimeDependentSchrodingerEquation}
i\,|\mathrm{\Psi}'(z)\rangle=i\,\frac{\partial}{\partial z}\,|\mathrm{\Psi}(z)\rangle=-\mathcal{Q}(z)\,|\mathrm{\Psi}(z)\rangle,
\end{equation}
here written in Dirac's notations, where the Hamilton-type operator $-\mathcal{Q}(z)$ is a function of the timelike coordinate $z$ and is supposed to be Hermitian. Denoting by $\{q_{m}(z)\}_{m}$ the set of the (real) eigenvalues of $\mathcal{Q}(z)$, assumed discrete, and by $\{|\psi_{m}(z)\rangle\}_{m}$ the one of the corresponding eigenvectors, assumed to constitute an orthonormal basis: $\langle\psi_{n}(z)|\psi_{n'}(z)\rangle=\delta_{n,n'}$, the solution of Equation~\eqref{Eq:TimeDependentSchrodingerEquation} may be generically expanded as
\begin{equation}
\label{Eq:EigenbasisExpansion}
|\mathrm{\Psi}(z)\rangle=\sideset{}{_{m}}\sum\alpha_{m}(z)\,e^{i\theta_{m}(z)}\,|\psi_{m}(z)\rangle,
\end{equation}
where
\begin{equation}
\label{Eq:DynamicalPhase}
\theta_{n}(z)=\int_{0}^{z}dz'\,q_{n}(z')
\end{equation}
denotes the ``dynamic'' phase of the propagating state $\alpha_{n}(z)\,e^{i\theta_{n}(z)}\,|\psi_{n}(z)\rangle$.

Substituting Equation~\eqref{Eq:EigenbasisExpansion} into Equation~\eqref{Eq:TimeDependentSchrodingerEquation}, one gets, making use of the eigenvalue equation $\mathcal{Q}(z)\,|\psi_{n}(z)\rangle=q_{n}(z)\,|\psi_{n}(z)\rangle$,
\begin{align}
\notag
&\left.\sideset{}{_{m}}\sum\alpha_{m}'(z)\,e^{i\theta_{m}(z)}\,|\psi_{m}^{\vphantom{\prime}}(z)\rangle\right. \\
\label{Eq:1stStepCalculation}
&\left.\quad=-\sideset{}{_{m}}\sum\alpha_{m}^{\vphantom{\prime}}(z)\,e^{i\theta_{m}(z)}\,|\psi_{m}'(z)\rangle,\right.
\end{align}
in such a way that, projecting onto the $n$th eigenstate $|\psi_{n}(z)\rangle$ of $\mathcal{Q}(z)$,
\begin{equation}
\label{Eq:2ndStepCalculation}
\alpha_{n}'(z)=-\langle\psi_{n}^{\vphantom{\prime}}(z)|\psi_{n}'(z)\rangle\,\alpha_{n}^{\vphantom{\prime}}(z)+\mathcal{R}_{n}^{\vphantom{\prime}}(z),
\end{equation}
where the rest
\begin{subequations}
\label{Eq:Rest}
\begin{align}
\notag
\mathcal{R}_{n}^{\vphantom{\prime}}(z)&\left.=-\sideset{}{_{m\neq n}}\sum\langle\psi_{n}^{\vphantom{\prime}}(z)|\psi_{m}'(z)\rangle\right. \\
\label{Eq:Rest-a}
&\left.\hphantom{=}\times e^{i[\theta_{m}(z)-\theta_{n}(z)]}\,\alpha_{m}(z)\right. \\
\notag
&\left.=-\sideset{}{_{m\neq n}}\sum\frac{\langle\psi_{n}(z)|\,\mathcal{Q}'(z)\,|\psi_{m}(z)\rangle}{q_{m}(z)-q_{n}(z)}\right. \\
\label{Eq:Rest-b}
&\left.\hphantom{=}\times e^{i[\theta_{m}(z)-\theta_{n}(z)]}\,\alpha_{m}(z).\right.
\end{align}
\end{subequations}
Equation \eqref{Eq:Rest-b} is obtained from the projection of the derivative with respect to $z$ of the eigenvalue equation $\mathcal{Q}(z)\,|\psi_{n'}(z)\rangle=q_{n'}(z)\,|\psi_{n'}(z)\rangle$ onto the $|\psi_{n}(z)\rangle$ eigenstate ($n\neq n'$) of $\mathcal{Q}(z)$:
\begin{align}
\notag
&\left.\langle\psi_{n}^{\vphantom{\prime}}(z)|\,\mathcal{Q}'(z)\,|\psi_{n'}^{\vphantom{\prime}}(z)\rangle+\langle\psi_{n}^{\vphantom{\prime}}(z)|\,\mathcal{Q}(z)\,|\psi_{n'}'(z)\rangle\right. \\
\label{Eq:3rdStepCalculation}
&\left.\quad=q_{n'}^{\vphantom{\prime}}(z)\,\langle\psi_{n}^{\vphantom{\prime}}(z)|\psi_{n'}'(z)\rangle,\right.
\end{align}
and from the identity
\begin{equation}
\label{Eq:4thStepCalculation}
\langle\psi_{n}^{\vphantom{\prime}}(z)|\,\mathcal{Q}(z)\,|\psi_{n'}'(z)\rangle=q_{n}^{\vphantom{\prime}}(z)\,\langle\psi_{n}^{\vphantom{\prime}}(z)|\psi_{n'}'(z)\rangle.
\end{equation}

In the particular case where $\mathcal{Q}(z)$ is an adiabatically varying function of $z$, the off-diagonal components of the rate of change of $\mathcal{Q}(z)$ in the $\{|\psi_{m}(z)\rangle\}_{m}$ eigenbasis and in units of the corresponding eigenvalue gap is small compared this gap \cite{Born1928, Messiah1999}, i.e.,
\begin{equation}
\label{Eq:ExactAdiabaticCondition}
\underset{z\in[0,L]}{\max}\,\bigg|\frac{\langle\psi_{n}(z)|\,\mathcal{Q}'(z)\,|\psi_{n'}(z)\rangle}{q_{n'}(z)-q_{n}(z)}\bigg|\ll\underset{z\in[0,L]}{\min}\,|q_{n'}(z)-q_{n}(z)|
\end{equation}
for all $(n,n'\neq n)$. Accordingly, $\mathcal{R}_{n}(z)$ in Equation~\eqref{Eq:2ndStepCalculation} may be neglected, yielding
\begin{equation}
\label{Eq:AdiabaticCoefficient}
\alpha_{n}(z)\simeq\alpha_{n}(0)\,e^{i\gamma_{n}(z)},
\end{equation}
where
\begin{equation}
\label{Eq:GeometricalPhase}
\gamma_{n}^{\vphantom{\prime}}(z)=i\int_{0}^{z}dz'\,\langle\psi_{n}^{\vphantom{\prime}}(z')|\psi_{n}'(z')\rangle.
\end{equation}
The latter is a real quantity because $\langle\psi_{n}^{\vphantom{\prime}}(z)|\psi_{n}'(z)\rangle$ is a purely imaginary number, as it can be demonstrated from the differenciation of the normalization condition $\langle\psi_{n}(z)|\psi_{n}(z)\rangle=1$. Inserting Equation~\eqref{Eq:AdiabaticCoefficient} into Equation~\eqref{Eq:EigenbasisExpansion}, one eventually obtains
\begin{equation}
\label{Eq:AdiabaticSolution}
|\mathrm{\Psi}(z)\rangle\simeq\sideset{}{_{m}}\sum\alpha_{m}(0)\,e^{i\theta_{m}(z)}\,e^{i\gamma_{m}(z)}\,|\psi_{m}(z)\rangle.
\end{equation}

As a result, in the case where the optical wave is initially in the $n$th eigenstate $|\psi_{n}(z)\rangle$, i.e., if $|\mathrm{\Psi}(0)\rangle=A_{n}\,|\psi_{n}(0)\rangle$, all the $\alpha_{m}(0)$'s in Equation~\eqref{Eq:AdiabaticSolution} are equal to $A_{m}\,\delta_{m,n}$ and the system then remains in the $|\psi_{n}(z)\rangle$ state:
\begin{equation}
\label{Eq:AdiabaticEigensolution}
|\mathrm{\Psi}(z)\rangle\simeq A_{n}\,e^{i\theta_{n}(z)}\,e^{i\gamma_{n}(z)}\,|\psi_{n}(z)\rangle,
\end{equation}
as it would do in the case of a $z$-independent process, only picking up a couple of phase factors in the course of the propagation along the $z$ axis. When the adiabatic effective evolution is not cyclic, i.e., when $\mathcal{Q}(L)\neq\mathcal{Q}(0)$, the phase factor $e^{i\gamma_{n}(z)}$ can be canceled out by a trivial choice of gauge for the eigenvectors, that is, $|\psi_{n}(z)\rangle\longmapsto|\tilde{\psi}_{n}(z)\rangle=e^{i\gamma_{n}(z)}\,|\psi_{n}(z)\rangle$. In the contrary case, $\gamma_{n}(z)$ becomes a gauge-invariant geometrical quantity known in quantum mechanics as the Berry phase \cite{Berry1988}.

\end{document}